\documentclass[journal]{new-aiaa}

\usepackage[utf8]{inputenc}
\usepackage{graphicx}
\usepackage{tikz}
\usepackage{amsmath}

\usepackage{algorithm}
\usepackage{algpseudocode}
\usepackage{mathtools}

\usepackage{amsthm}

\newcommand{\opnorm}[1]{\left\lvert\hspace{-1 pt}\left\lvert\hspace{-1 pt}\left\lvert#1\right\lvert\hspace{-1 pt}\right\lvert\hspace{-1 pt}\right\lvert}

\makeatletter
\newcommand{\onetagright}{\tagsleft@false}
\makeatother

\usepackage[version=4]{mhchem}
\usepackage{siunitx}
\usepackage{longtable,tabularx}
\title{Satellite Safe Margin: Fast Solutions for Conjunction Analysis}

\author{Ricardo N. Ferreira \footnote{Corresponding author} \footnote{PhD Student, Department of Computer Science, rjn.ferreira@campus.fct.unl.pt}}
\affil{NOVA School of Science and Technology, 2829-516, Caparica, Portugal}
\author{Marta Guimar\~{a}es \footnote{AI Research Lead, AI Department, marta.guimaraes@neuraspace.com}}
\affil{Neuraspace, 3030-199, Coimbra, Portugal}
\author{Cláudia Soares \footnote{Assistant Professor, Department of Computer Science, claudia.soares@fct.unl.pt}}
\affil{NOVA School of Science and Technology, 2829-516, Caparica, Portugal}

\begin{document}

\maketitle

\begin{abstract}
The amount of debris in orbit has increased significantly over the years. With the recent growth of interest in space exploration, conjunction assessment has become a central issue.
One important metric to evaluate conjunction risk is the miss distance. However, this metric does not intrinsically take into account uncertainty distributions. Some work has been developed to consider the uncertainty associated with the position of the orbiting objects, in particular, to know if these uncertainty distributions overlap (e.g., ellipsoids when considering Gaussian distributions). This work presents two fast solutions to not only check if the ellipsoids overlap but to compute the distance between them, referred to as margin.
One solution computes the margin when the best-known data from both objects can be centralized (e.g., debris-satellite conjunctions). The second solution is applied in the decentralized case when the most precise covariances cannot be shared (conjunctions of satellites owned by different operators). These methods are both accurate and fast, being able to process 15,000 conjunctions per minute with the centralized solution and approximately 490 conjunctions per minute with the distributed solution.
\end{abstract}

\section{Introduction}

\lettrine{T}{he} number of objects in orbit increases significantly each year. Objects in orbit range from operating satellites to pieces of broken satellites to tiny fragments like scraps of paint. This continuous increase in the number of objects in orbit, frequently denominated as the Kessler effect~\cite{kessler1978collision}, presents a danger to the integrity of functional satellites as well as to the future accessibility of humankind to space~\cite{pelton2013space,manfletti2023ai}. 

It is estimated that, in orbit, there are millions of fragments a few millimeters in size, several thousand a few centimeters in size, and a few thousand inoperative satellites and discarded rocket stages. Due to the high speeds that these fragments can reach, in the
order of thousands of kilometers per hour, they can cause fractures in a satellite’s hull, for example. An object larger than 10 cm in
size, traveling at such speeds could make communications or other important
satellite resources unusable \cite{pelton2013space}. This evolution leads to an increase in the attention of the scientific community to various problems, such as collision avoidance between satellites \citep{kleinig2022collision,bonnal2020just,mishne2017collision,reiland2021assessing} and collision risk \citep{le2018space,lucken2019collision,balch2019satellite}.

Tracking and monitoring of objects in orbit is associated with uncertainty \cite{poore2016covariance}, from which some types stand out: Structural uncertainty (or model bias); Sensor measurement noise; Propagation of uncertainty; and Algorithmic uncertainty. All these factors have a significant impact on conjunction assessment through the probability of collision, satellite evasion maneuvers or simply tracking the trajectories of objects in orbit \cite{poore2016covariance,klinkrad2006space}. Hereupon, it becomes clear the large number of threats to the integrity of the various assets in orbit. These satellites play a crucial role in numerous services on the planet, such as communications, Earth observation, scientific experiments, and weather monitoring. Therefore, it is important to preserve the good functioning of these satellites, thus avoiding the collision of space debris with active spacecraft.

Nowadays, a very important metric in the evaluation of conjunctions is the \emph{miss
distance}, which corresponds to the distance between the two objects at
the closest point of approach \cite{klinkrad2006space}. Another important tool is the \emph{probability of collision}, which reflects the risk of collision considering the miss distance and the uncertainty associated with the positions of the objects at the Time of Closest Approach (TCA)~\cite{klinkrad2005collision}. These two tools allow us to evaluate possible collisions in the future and plan evasion maneuvers to avoid them. Cases of concern are classified as encounters whose Time of Closest Approach (TCA) occurs within five days and whose probability of collision is greater than $10^{-7}$~\cite{nasahandbook}. Cases of concern will be followed up to the TCA and those involving active satellites result in a notification for the satellite Owner/Operator (O/O) of a probable conjunction~\cite{nasahandbook}.

From an operational perspective, it is often assumed that the uncertainty associated with the positions of objects follows a normal distribution \cite{merzesa,poore2016covariance}. Thus, we can view the positions of objects as random variables whose mean value is the measured position value and, since they follow a Gaussian distribution, in a three-dimensional space, the uncertainty can be seen as an ellipse. In this setting, the \emph{miss distance} is the difference between the mean vectors of two probability distributions of object positions. 

Some work has been developed to explore how we can exploit the information of the uncertainty associated with the position of orbiting objects to enhance the analysis of satellite conjunctions. Over the years, different solutions have been presented to bound the probability of collision.~\citeauthor{alfriend1999probability} present a maximum value for the probability of collision for a particular conjunction~\cite{alfriend1999probability}.~\citeauthor{ferreira2023probability} present tight and fast-to-compute upper and lower bounds for the probability of collision~\cite{ferreira2023probability}. Another avenue of research explores the identification of cases of concern by checking if the uncertainty ellipsoids of the objects overlap~\cite{alfano2003determining}. However, some available solutions do not consider the size of the objects, which for small position uncertainties should not be neglected~\cite{balch2019satellite}.

Following the idea of ellipses overlap for conjunction analysis, we present fast solutions to not only check if the ellipsoids overlap but to compute the distance between them, which we call \emph{margin}. This allows an operator to have access to new information: for a chosen $k$-$\sigma$ uncertainty ellipsoid, an operator has the guarantee that the true distance between the two objects is not smaller than the computed \emph{margin}, which by definition is the smallest possible distance compatible with an operator risk aversion (assuming that the true position of the objects lies in the chosen uncertainty ellipsoid).~\footnote{The \emph{margin} is not a metric but instead a pseudo-semi-metric. It is a pseudo-metric since it does not respect the axiom $d(x,y) = 0 \iff x = y$, as we can have two different ellipsoids $x$ and $y$ whose distance between them is zero. The \emph{margin} is at the same time semi-metric as it does not necessarily satisfy the triangle inequality. We can have three ellipsoids $x$, $y$ and $z$ such that $d(x,y) = d(y,z) = 0$ and $d(x,z) \neq 0$.}

\begin{figure}[ht!]
    \centering
    \includegraphics[scale=0.4]{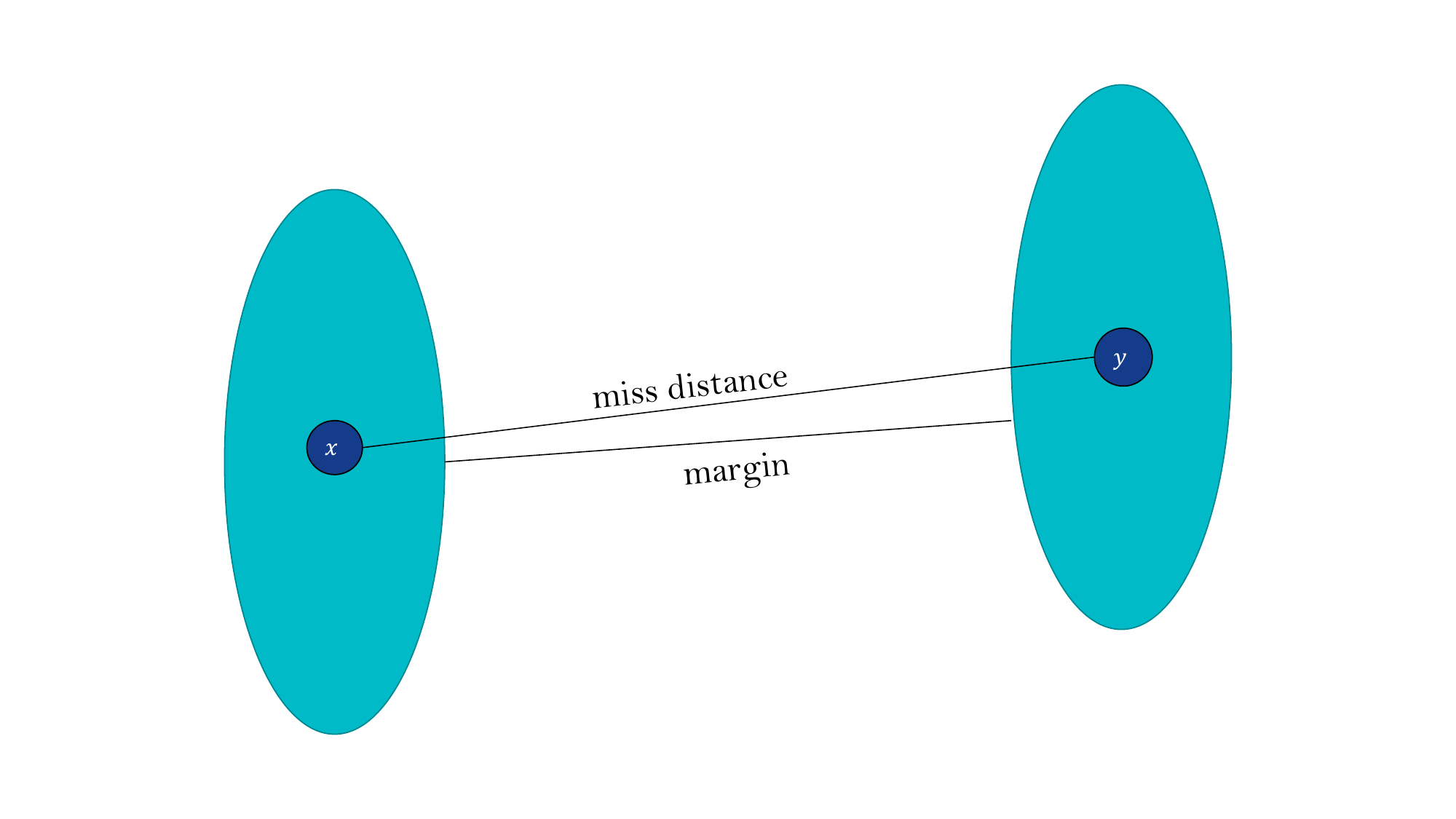}
    \includegraphics[scale=0.4]{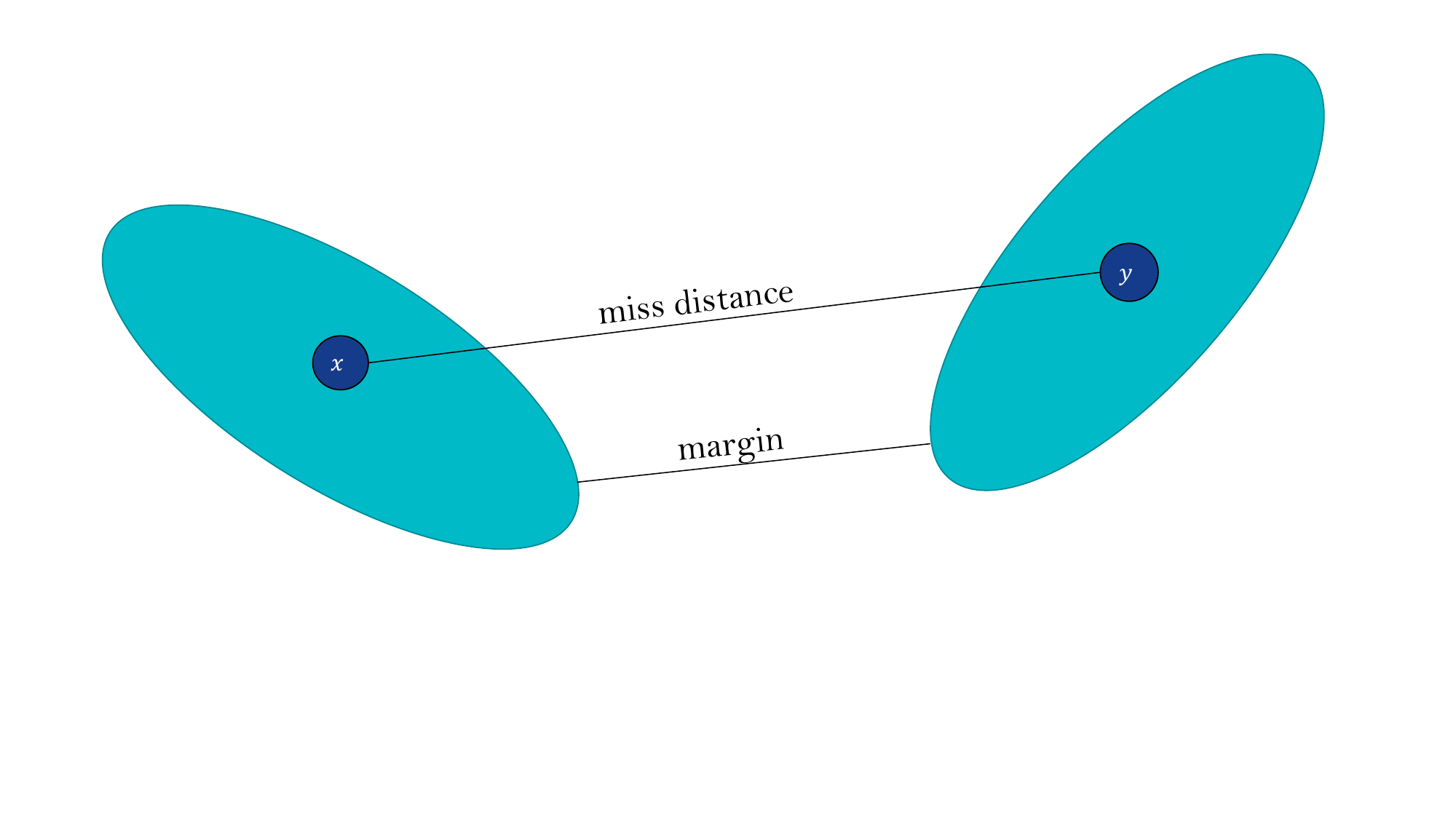}
    \caption[Representation of the difference between the miss distance and the margin in two different conjunctions.]{Representation of the difference between the miss distance and the margin in two different conjunctions.} 
    \label{fig:margin-two-representations}
\end{figure}
In Fig.~\ref{fig:margin-two-representations}, we have a representation contrasting the properties of the miss distance and the margin in two different conjunctions. In both cases, we have the same miss distance. However, due to the orientation of the uncertainty ellipsoids, the margin in the case above is greater than the margin in the conjunction below. Analyzing these cases using the classic miss distance, an operator would not see the difference. However, the second case is more concerning as for an operator-specified uncertainty ellipsoid, assuming that it contains the true position of the objects, the smallest possible distance is considerably smaller.

While not directly linked, this problem resembles the problem of collision detection~\cite{tracy2023differentiable,rimon1997obstacle,canny1986collision}, an important problem with applications in robotics and computer graphics. The goal is to assess if two objects are in contact with each other at one or more points. For simplicity, a large part of the literature focuses on convex primitives (e.g., polytopes, ellipsoids, cylinders, cones), as we can approximate any shape with its convex hull~\cite{kockara2007collision}. The first algorithm developed for collision detection, the Gilbert-Johnson-Keerthi (GJK) algorithm~\cite{gilbert1988fast}, computes the distance and closest points between the convex hulls of two sets of vertices.~\citeauthor{rimon1997obstacle} simplify the shape of
objects by resorting to minimum-volume enclosing ellipsoids and perform an eigenvalue-based technique to estimate
the distance between the ellipses~\cite{rimon1997obstacle}. One research focus is the collision detection of moving objects~\cite{canny1986collision,tracy2023differentiable}. For example,~\citeauthor{tracy2023differentiable} present differentiable collision detection for a set of different convex primitives and use their result for trajectory optimization of a quadrotor~\cite{tracy2023differentiable}. Recently, first-order methods, such as the Frank-Wolfe algorithm and the projected gradient descent method, and acceleration techniques, such as Polyak acceleration~\cite{polyak1964some} and Nesterov acceleration~\cite{nesterov1983method}, have been explored to speed up the computation needed to solve collision detection problems~\cite{montaut2024gjk++}.

Our problem, in particular, is not the space object but instead the uncertainty that is seen as a convex primitive, the ellipsoid. In this work, we present two fast solutions to compute the \emph{margin}, which represents the smallest possible distance between two objects considering the uncertainty associated with their positions and assuming that the true position of the objects lies on their uncertainty ellipsoids.

Recently, first-order methods have regained relevance due to the low computational cost and have been significant advancements in acceleration techniques for convex optimization. Following the good results of these techniques for collision detection, we resort to the Frank-Wolfe algorithm and the Fast Iterative Shrinkage-Thresholding, each applied in a different case:

\begin{itemize}
    \item When the best-known data from both objects can be centralized (e.g. debris-satellite conjunctions or conjunctions between satellites from the same operator), we resort to the Frank-Wolfe Algorithm;
    \item  When the most precise covariances cannot be shared (e.g., conjunctions of satellites owned by different operators), we resort to the Fast Iterative Shrinkage-Thresholding Algorithm.
\end{itemize}

For circumstances in which sharing information is not desirable, we use the Fast Iterative Shrinkage-Thresholding Algorithm (FISTA). Due to the structure of our problem, we can use FISTA to solve the problem in a distributed way. This last option allows for collaboration between satellite operators when more precise information about assets is not to be shared directly while not neglecting accuracy. Both methods outperform, in accuracy, our benchmark for computing distances between ellipses.


In Section~\ref{sec:problem-statement}, we present and describe the problem we intend to solve. We define the computation of the margin as a convex optimization problem whose goal is to minimize the distance between two objects considering the uncertainty associated with their positions. In Section-\ref{sec:rimon-boyd}, as a benchmark, we present a method proposed by~\citeauthor{rimon1997obstacle} to calculate the distance between two ellipses. However, this approach is extremely sensitive to numerical errors due to the diagonalization of non-normal matrices, thus not being acceptable for this context in which operators must make high-stakes decisions. So, in the subsequent sections, we present our two solutions to compute the margin. In Section~\ref{sec:frank-wolfe}, we introduce the Frank-Wolfe algorithm and its implementation for our specific problem. In Section~\ref{sec:fista}, we present the Fast Iterative Shrinkage-Thresholding Algorithm and explain the implementation to compute the margin, jointly with a fast method for the projection step. Lastly, in Section~\ref{sec:experiments}, we show some numerical experiments, comparing the accuracy and processing times of these three methods, considering a ground truth computed by the general solver CVXPY~\cite{diamond2016cvxpy,agrawal2018rewriting}. We also present preliminary results that motivate the use of the margin to aid in the decision-making process for conjunction analysis. At the end of the section, we discuss how this pseudo-semi-metric compares with other tools such as the Mahalanobis distance~\cite{mahalanobis1936generalised} and the Kullback-Leibler divergence~\cite{kullback1951information} as well as the applicability for operators.
\section{Problem Statement}
\label{sec:problem-statement}

In this section, we concretely define the problem we intend to solve, which will serve as a basis for the implementation of the various methods presented in the next sections.

From an operational perspective, it is considered that the uncertainty associated with the positions of objects follows a normal distribution \cite{merzesa,poore2016covariance}. Thus, we can view the positions of objects as random variables whose mean value is the measured position value and, in a three-dimensional space, the distribution described by the covariance can be seen as an ellipse. The \emph{miss distance} is the difference between the mean values. In this work, we intend to accurately and expeditiously compute the distance between the uncertainty ellipsoids. The idea behind this is: assuming that the true position is in the error ellipse, in a worst-case perspective, what would be the smallest possible distance between the objects?

Let $x_0$ be the random variable of the chaser position and $y_0$ be the random variable of the target position. We assume that the uncertainties in the positions are independent and defined by Gaussian distributions. So, we can define the random variables as $x_0 \sim \mathcal{N}\left(\overline{x}_0, \Sigma_{x_0} \right)$ and $y_0 \sim \mathcal{N}\left(\overline{y}_0, \Sigma_{y_0} \right)$, such that $\overline{x}_0$ and $\overline{y}_0$ are the positions of the chaser and the target, respectively, with $\Sigma_{x_0}$ and $\Sigma_{y_0}$ being the corresponding covariance matrices. Since we consider that the random variables follow a Gaussian distribution, their equiprobable lines can be seen as ellipses, such that the true position of the object lies in the ellipse with some probability. Thus, we can define the error ellipses of the two objects as

\begin{equation}
  \begin{aligned}
    &\xi_x = \left\{ p \in \mathbb{R}^3 : (p - \overline{x}_0)^T \Sigma_{x_0}^{-1} (p - \overline{x}_0) \leq 1 \right\} \\
    &\xi_y = \left\{ p \in \mathbb{R}^3 : (p - \overline{y}_0)^T \Sigma_{y_0}^{-1} (p - \overline{y}_0) \leq 1 \right\}.
  \end{aligned}
\end{equation}

It is worth highlighting that this formulation is equivalent to any level of uncertainty. We can divide the matrices $\Sigma_{x_0}^{-1}$ and $\Sigma_{y_0}^{-1}$ by a value $\sigma$ that quantifies how much probability is contained in the uncertainty ellipsoids.

\begin{figure}[h!]
    \centering
    \includegraphics[scale=0.9]{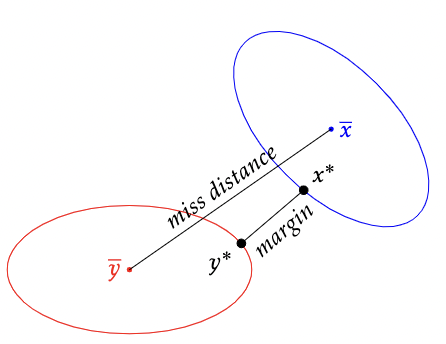}
    \caption{Two-dimensional representation of the $x^*$ and $y^*$ points for the margin computation between ellipses.}
    \label{fig:margin-representation}
\end{figure}

Assuming that the true positions are on the ellipses, in a worst-case perspective, we want to know the minimum relative distance that the two objects can have between them. In other words, we want to calculate the possible minimum distance between the ellipses. Our problem can be formulated as the convex optimization problem

\begin{equation}
    \begin{aligned}
        &\underset{x, y \ \in \ \mathbb{R}^{3}}{\mbox{minimize}} \quad \| x - y \|^2 \\
        &\mbox{subject to} \\ 
        &\quad \quad (x - \overline{x}_0)^T \Sigma_{x_0}^{-1} (x - \overline{x}_0) \leq 1 \\
        &\quad \quad (y - \overline{y}_0)^T \Sigma_{y_0}^{-1} (y - \overline{y}_0) \leq 1
    \end{aligned}
    \label{eq:margin-optimization-problem}
\end{equation}
In Fig.~\ref{fig:margin-representation}, we depict our problem. Assuming that the actual positions of the objects lie on their uncertainty ellipsoids, we want to find the possible positions of the objects, $x^*$ and $y^*$, which are closest to each other. As we can see, the \emph{margin} is always smaller than the \emph{miss distance} which would be the distance between the mean values $\overline{x}$ and $\overline{y}$. This safety distance can help to assess possible risk situations that will go undetected by considering only the miss distance.

In the following sections, we present the analytical method proposed by \citeauthor{rimon1997obstacle}. However, it gives wrong results in some cases due to numerical errors in the eigenvalue decomposition of a non-diagonalizable matrix. To overcome these problems, we present two ways to calculate the \emph{margin} accurately, one in a centralized way, using the Frank-Wolfe Algorithm, and the other in a distributed way, using the Fast Iterative Shrinkage-Thresholding Algorithm. This last option allows for collaboration between satellite operators when more precise information about assets is not to be shared directly while not neglecting accuracy.
\section{Setting the Benchmark: the Rimon-Boyd Approach}
\label{sec:rimon-boyd}

We now present the approach proposed by Rimon and Boyd for computing the margin between ellipses \cite{rimon1997obstacle}. The aim of the method by \citeauthor{rimon1997obstacle} is to detect obstacle collisions. The authors simplify the shape of
objects by resorting to minimum-volume enclosing ellipsoids and showing that the distance between ellipsoids can be seen as an eigenvalue problem.

This method applies to mechanical objects whose initial positions guarantee that the ellipses enclosing the objects do not intercept. However, in our specific case, the Conjunction Data Messages (CDMs) return information on the positions of the objects and the associated uncertainty ellipsoid at the Time of Closest Approach (TCA). In this setting, there is no guarantee that the ellipses do not intercept. Therefore, we must first check if the ellipses intercept; otherwise, we can apply this method to compute the distance between the ellipses. Our implementation consists of two steps: first, we check if the ellipses overlap, using a simple condition for overlap of $n$-dimensional ellipses presented by Gilitschenski and Hanebeck \cite{gilitschenski2012robust}, which consists of evaluating a convex scalar function. If they do not overlap, we apply the algorithm for computing the margin between ellipses presented by \citeauthor{rimon1997obstacle}. If the ellipses overlap, then the value of \emph{margin}, the minimum distance between the ellipses, is zero. From this point on, let us consider our error ellipses defined in~\eqref{eq:margin-optimization-problem} as

\begin{equation}
  \label{eq:margin-error-ellipses}
  \begin{aligned}
    &\xi_x = \left\{ p \in \mathbb{R}^3 : (p - b)^T B (p - b) \leq 1 \right\} \\
    &\xi_y = \left\{ p \in \mathbb{R}^3 : (p - c)^T C (p - c) \leq 1 \right\} ,
  \end{aligned}
\end{equation}
such that $b = \overline{x}_0$, $B = \Sigma_x^{-1}$, $c = \overline{y}_0$ and $C = \Sigma_y^{-1}$.

\subsection{Ellipse Overlap Test}

\noindent Every point $p$ belonging to the intersection of the ellipses, $p \in \xi_x \cap \xi_y$, satisfies

\begin{equation}
  \label{eq:overlap-condition}
  \begin{aligned}
    \lambda \ (p - b)^T B \, (p - b) + (1 - \lambda) \ (p - c)^T C \, (p - c) \leq 1 \ , \quad \quad \mbox{for} \ \lambda \in [0,1].
  \end{aligned}
\end{equation}
In fact, this expression describes the union of the two ellipses, $\xi_x \cup \xi_y$. For $\lambda = 0$, we obtain $\xi_x$; for $\lambda = 1$, we obtain $\xi_y$. Following the first proposition presented by the authors, we have that the set of points that satisfies~\eqref{eq:overlap-condition} for $\lambda \in [0,1]$ is empty, one single point or an ellipsoid $\xi_{\lambda} = \left\{ p \in \mathbb{R}^n : (p - m_{\lambda})^T \hat{\Sigma}_{\lambda}^{-1} (p - m_{\lambda}) \leq 1 \right\}$, where

\begin{equation}
  \label{eq:overlap-ellipse-variables}
  \begin{aligned}
    &\hat{\Sigma}_{\lambda} = \frac{\Sigma_{\lambda}}{K(\lambda)} \\
    &\Sigma_{\lambda} = \lambda \ B + (1 - \lambda) \ C  \\
    &m_{\lambda} = \Sigma_{\lambda}^{-1} (\lambda \ B \, b + (1 - \lambda) \ C \, c) \\
    &K(\lambda) = 1 - \lambda \ b^T B \, b - (1 - \lambda) \ c^T C \, c + m_{\lambda}^T \Sigma_{\lambda} m_{\lambda}
  \end{aligned}
\end{equation}
Using these variables, we can transform~\eqref{eq:overlap-condition} into

\begin{equation}
  \label{eq:transformed-overlap-condition}
  \begin{aligned}
    (p - m_{\lambda})^T \Sigma_{\lambda} \, (p - m_{\lambda}) \leq K(\lambda).
  \end{aligned}
\end{equation}
For $K(\lambda) > 0$, the expression~\eqref{eq:transformed-overlap-condition}, dividing by $K(\lambda)$, describes the ellipsoid $\xi_{\lambda}$; for $K(\lambda) = 0$, the set of points that satisfies the expression is $m_{\lambda}$; and for $K(\lambda) < 0$, the division of~\eqref{eq:transformed-overlap-condition} by $K(\lambda)$ results in an inequality, which is not satisfied by any point $p$.

In the second proposition, the authors prove that the two ellipses do not overlap if and only if there exists $\lambda^* \in [0,1]$ with $K(\lambda^*) < 0$. Thus, if the minimum of the function $K$ is less than zero, then we can conclude that the ellipses do not overlap; otherwise, they overlap. The function $K: [0,1] \mapsto \mathbb{R}$ is a convex scalar function of $\lambda$ that can also be represented as

\begin{equation}
  \label{eq:K-second-representation}
  \begin{aligned}
    K(\lambda) = 1 - \lambda \ (1 - \lambda) \ (c - b)^T C \Sigma_{\lambda}^{-1} B (c - b) \ .
  \end{aligned}
\end{equation}
Since the function $K$ is convex, we can apply Brent's method \cite{brent1971algorithm} to calculate the minimum of the function, given that it is computationally fast and whose implementations are available in various libraries, such as Scipy \cite{2020SciPy-NMeth}.  

\subsection{Margin Computation}

Now that we can rapidly check if the two ellipses overlap or not, we can advance to the computation of the margin (the distance between the ellipsoids) presented by~\citeauthor{rimon1997obstacle}. For simplification, we will maintain the same notation that was used in the original paper.

In summary, to compute the distance estimate between our two ellipsoids, $\xi_x$ and $\xi_y$, we start by computing $y^* \in \xi_y$ at which the ellipsoidal level surfaces surrounding $\xi_x$ first touch $\xi_y$. Then we compute $x^* \in \xi_x$ which is the closest to $y^*$. The distance estimate is then margin($\xi_x$, $\xi_y$) = $\| x^* - y^* \|$. The authors summarize this computation in the following proposition:

\textbf{Proposition 1.}
\label{prop:margin-estimation}
Given two ellipsoids $\xi_x$ and $\xi_y$, the point $y^* \in \xi_y$ is $y^* = b + \lambda_1 B^{-\frac{1}{2}} [\lambda_1 I - \Tilde{C}]^{-1} \bar{c}$, where $\Tilde{C} = \overline{C} \,^{-1}$ and $\lambda_1$ is the minimal eigenvalue of the $2n \times 2n$ matrix

\begin{equation}
    \label{eq:min-eigenvalue-1}
    \begin{bmatrix}
            \Tilde{C} & -I \\
            -\Tilde{c}\Tilde{c}^T & \Tilde{C}
        \end{bmatrix},
\end{equation}
such that $\overline{C} = B^{-\frac{1}{2}} C B^{-\frac{1}{2}}$ and $\Tilde{c} = \overline{C} \,^{-\frac{1}{2}} B^{\frac{1}{2}}(c - b)$. The point $x^* \in \xi_x$ is $x^* = y^* + \mu_1 [\mu_1 I - \Tilde{B}]^{-1} b$, where $\Tilde{B} = B^{-1}$ and $\mu_1$ is the minimal eigenvalue of the $2n \times 2n$ matrix

\begin{equation}
    \label{eq:min-eigenvalue-2}
    \begin{bmatrix}
            \Tilde{B} & -I \\
            -\Tilde{b}\Tilde{b}^T & \Tilde{B}
        \end{bmatrix},
\end{equation}
such that $\Tilde{b} = \overline{B} \,^{-\frac{1}{2}} b$. The distance estimate margin($\xi_x$, $\xi_y$) = $\| x^* - y^* \|$, is positive when $\xi_x$ and $\xi_y$ are disjoint, and zero when $\xi_x$ and $\xi_y$ touch such that their interiors are disjoint.

All the details of this proposition can be found in the original paper~\cite{rimon1997obstacle}. However, in some cases, this method gives wrong results, as we will see when comparing the results of the various methods. This error in the results happens because the matrix for which we want to find the smallest eigenvalue is not normal. Let $M$ be the matrix presented in~\eqref{eq:min-eigenvalue-1}, to demonstrate what has been stated. The real matrix $M$ is normal if $M^T M = M M^T$. It is known that the eigenvalues of a non-normal matrix $M$ are not indicative of the behavior of the operator applied to the relevant vector space. 

\begin{equation}
    \label{eq:M-diagonalization}
    \begin{aligned}
    M^T M &= M M^T \\
    \begin{bmatrix}
        \Tilde{C} & -\Tilde{c}\Tilde{c}^T \\
        -I & \Tilde{C}
    \end{bmatrix} \begin{bmatrix}
        \Tilde{C} & -I \\
        -\Tilde{c}\Tilde{c}^T & \Tilde{C}
    \end{bmatrix} &= \begin{bmatrix}
        \Tilde{C} & -I \\
        -\Tilde{c}\Tilde{c}^T & \Tilde{C}
    \end{bmatrix} \begin{bmatrix}
        \Tilde{C} & -\Tilde{c}\Tilde{c}^T \\
        -I & \Tilde{C}
    \end{bmatrix} \\
    \begin{bmatrix}
        \Tilde{C}^2 + (-\Tilde{c}\Tilde{c}^T)^2 & -\Tilde{C} + (-\Tilde{c}\Tilde{c}^T) \Tilde{C} \\
        -\Tilde{C} + (-\Tilde{c}\Tilde{c}^T) \Tilde{C} & \Tilde{C}^2 + I^2
    \end{bmatrix} &= \begin{bmatrix}
        \Tilde{C}^2 + I^2 & -\Tilde{C} + (-\Tilde{c}\Tilde{c}^T) \Tilde{C} \\
        -\Tilde{C} + (-\Tilde{c}\Tilde{c}^T) \Tilde{C} & \Tilde{C}^2 + (-\Tilde{c}\Tilde{c}^T)^2
    \end{bmatrix}.
    \end{aligned}
\end{equation}
Therefore, for M to be normal, we have that $(-\Tilde{c}\Tilde{c}^T)^2$ has to be equal to $I^2$. However, it is quickly seen that this is not possible since $-\Tilde{c}\Tilde{c}^T$ is a matrix of rank one while $I$ is a matrix of rank three. Therefore, $M$ is not normal. This is a problem because non-normal matrices are extremely sensitive to numerical errors in the computation of their eigenvalues and these may not be accurate. An inaccurate eigenvalue implies a wrong result in the estimation of the distance between ellipses since the result depends exclusively on the computation of the smallest eigenvalue of the matrix M.

\section{Optimal and Stable solution with data sharing: Iterative Frank-Wolfe algorithm}
\label{sec:frank-wolfe}

To overcome the problems raised in the previous section, we can resort to first-order iterative methods, which provide simple and computationally cheap operations exploring only the objective function and its gradient. These simple operations avoid running into numerical and/or round-off problems.

When it is possible to centrally process the information of the objects involved in a conjunction (for example, debris-satellite conjunctions), we can use the Frank-Wolfe Algorithm to compute the \emph{margin} in a centralized way. In this section, we will analyze the two variants of the Frank-Wolfe Algorithm, seen earlier, for our specific problem.

\subsection{Frank-Wolfe Algorithm}

\noindent The Frank-Wolfe (FW) Algorithm \cite{frank1956algorithm} is a first-order method that aims to solve the general optimization problem

\begin{equation}
    \label{eq:frank-wolfe-problem}
    \begin{aligned}
        &\underset{x \in \mathcal{D}}{\mbox{minimize}} \quad \Psi(x)
    \end{aligned}
\end{equation}
such that $\mathcal{D}$ is a compact and convex set, and $\Psi: \mathcal{D} \rightarrow \mathbb{R}$ is a convex and differentiable function. At each iteration $k$, the algorithm aims to minimize a linear approximation $\phi(s^k) = {s^k}^T \, \nabla \Psi( x^k )$ around the current iteration $x^k$ of the objective function over the same domain $\mathcal{D}$ of the original problem. In the constrained optimization setting, this method stands out for not needing a projection or proximal step to remain feasible.

The original Frank-Wolfe algorithm is systematized in Algorithm~\ref{alg:line-search-frank-wolfe}. At each iteration, after the minimization of the linear approximation, the next iteration moves in the direction of the computed minimizer with some step size $\alpha$ at the $k$-th iteration.

The algorithm has a convergence rate for the objective of $O\left(\frac{1}{k}\right)$ and has regained relevance in several areas given its low implementation cost and good scalability. Based on the original algorithm, new variants of this algorithm have been presented and analyzed with relevant features to increase the progress in each iteration \cite{jaggi2013revisiting,pedregosa2020linearly,lacoste2015global,dvurechensky2022generalized,liu2016efficient,guelat1986some}. We will use one of these variants to solve our problem. Instead of using the predefined value of $\alpha = \frac{2}{k+2}$, we can find the best step size for each particular iteration through line search. So $\alpha$ can be chosen as

\begin{equation}
    \label{eq:frank-wolfe-line-search-alpha}
    \begin{aligned}
        \alpha = \underset{\alpha \in [0,1]}{\arg\min} \ \Psi(x^k + \alpha (s^k - x^k)).
    \end{aligned}
\end{equation}
Therefore, the next iteration is the best point belonging to the line segment between the current iteration $x^k$ and $s^k$.

\begin{algorithm}[ht!]
\caption{Frank-Wolfe with Line Search for step size $\alpha$}
\label{alg:line-search-frank-wolfe}
\begin{algorithmic}
\Require{$x \in \mathcal{D}$}
\State $x^0 \gets x$

\For{$k = 0,...,n$}
    \State $s^k \gets \underset{s \in \mathcal{D}}{\arg\min} \ s^T \nabla \Psi(x^k)$
    \State $\alpha \gets \underset{\alpha \in [0,1]}{\arg\min} \ \Psi(x^k + \alpha (s^k - x^k))$
    \State $x^{k+1} \gets x^k + \alpha (s^k - x^k)$
\EndFor
\end{algorithmic}
\end{algorithm}

\subsection{Optimal and stable margin computation}

For our specific case, as stated before, we are trying to solve Problem~\eqref{eq:margin-optimization-problem}. So, we are trying to minimize the objective function $f(x,y) = \|x - y\|^2$, such that $x \in \xi_x \coloneqq \left\{ p \in \mathbb{R}^3 : (p - \overline{x}_0)^T \Sigma_{x_0}^{-1} (p - \overline{x}_0) \leq 1 \right\}$ and $y \in \xi_y \coloneqq \left\{ p \in \mathbb{R}^3 : (p - \overline{y}_0)^T \Sigma_{y_0}^{-1} (p - \overline{y}_0) \leq 1 \right\}$, the uncertainty ellipsoids of object positions. Therefore, at each iteration, the first step of the algorithm consists of minimizing a linear approximation of the objective function over the same domain, i.e., consists of solving the following convex optimization problem

\begin{equation}
    \label{eq:frank-wolfe-first-step-problem}
    \begin{aligned}
        \underset{s_1, s_2 \in \mathbb{R}^3}{\mbox{minimize}} & \quad g\left( s_1, s_2 \right) = 2 \begin{bmatrix}
            s_1^T & s_2^T
        \end{bmatrix} \begin{bmatrix}
            x - y \\
            y - x
        \end{bmatrix} \\
        \mbox{subject to} & \quad \left\| s_1 - \overline{x}_0 \right\|^{2}_{\Sigma_{x_0}^{-1}} \leq 1 \\
        & \quad \left\| s_2 - \overline{y}_0 \right\|^{2}_{\Sigma_{y_0}^{-1}} \leq 1
    \end{aligned}
\end{equation}
where $g(\cdot)$ is the linear approximation around the current iteration of the objective function $f$, $\left\| u \right\|_{\Sigma} \coloneqq \sqrt{u^T \Sigma u}$ is the norm induced by the positive definite operator $\Sigma$. So, let us analyze how to compute the optimal solution $s^*$ for the minimization of the linear approximation of the objective function.

\textbf{Proposition 2.} 
At each iteration $k$, the optimal solution $s^* \coloneqq \left( s_1^*, s_2^* \right)$ for the minimization of the linear approximation of the objective function in~\eqref{eq:margin-optimization-problem} is given by

\begin{equation}
\begin{aligned}
    s_1^* &= \overline{x}_0 + \frac{\Sigma_{x_0} (y - x)}{\left\| y - x \right\|_{\Sigma_{x_0}}} \\
    s_2^* &= \overline{y}_0 + \frac{\Sigma_{y_0} (x - y)}{\left\| x - y \right\|_{\Sigma_{y_0}}} .
\end{aligned}
\end{equation}

\begin{proof}
The first step of the algorithm consists of solving the convex optimization problem in~\eqref{eq:frank-wolfe-first-step-problem}. By the Karush–Kuhn–Tucker (KKT) conditions~\cite{kuhn1951nonlinear,karush1939minima,bazaraa2006nonlinear}, the primal and dual optimal solutions respect the equations

\begin{equation}
    \label{eq:KKT-conditions}
    \begin{aligned}
        &\nabla g(s^*) + \lambda_1^* \nabla f_1(s^*) + \lambda_2^* \nabla f_2(s^*) = 0, & \\
        &\lambda_{i}^{*} f_i(s^*) = 0, &\quad \mbox{for } i = 1, 2, \\
        &\lambda_{i}^{*} \geq 0, &\quad \mbox{for } i = 1, 2. \\
    \end{aligned}
\end{equation}
where $f_1(s) \coloneqq \left\| s_1 - \overline{x}_0 \right\|^{2}_{\Sigma_{x_0}^{-1}} - 1 $ and $f_2(s) \coloneqq \left\| s_2 - \overline{y}_0 \right\|^{2}_{\Sigma_{y_0}^{-1}} - 1$ are the constraints of Problem~\eqref{eq:frank-wolfe-first-step-problem}, such that $f_1(s) \leq 0$ and $f_2(s) \leq 0$, and $g(\cdot)$ is the function we are trying to minimize.
Solving the first equation for the optimal primal solution $s^*$, we obtain

\begin{equation}
    \label{eq:first-KKT-condition}
    \begin{aligned}
        & 2 \begin{bmatrix}
            x - y \\
            y - x
        \end{bmatrix} + 2 \lambda_{1}^{*} \begin{bmatrix}
            \Sigma_{x_0}^{-1} (s_1^* - \overline{x}_0) \\
            0
        \end{bmatrix} + 2 \lambda_{2}^{*} \begin{bmatrix}
            0 \\
            \Sigma_{y_0}^{-1} (s_2^* - \overline{y}_0)
        \end{bmatrix} = 0 \iff \\ \\
        & \begin{bmatrix}
            \lambda_1^* \Sigma_{x_0}^{-1} & 0 \\
            0 & \lambda_2^* \Sigma_{y_0}^{-1}
        \end{bmatrix} \left( \begin{bmatrix}
            s_1^* \\
            s_2^*
        \end{bmatrix} - \begin{bmatrix}
            \overline{x}_0 \\
            \overline{y}_0
        \end{bmatrix} \right) = \begin{bmatrix}
            y - x \\
            x - y
        \end{bmatrix} \iff \\ \\
        & \begin{bmatrix}
            s_1^* \\
            s_2^*
        \end{bmatrix} = \begin{bmatrix}
            \overline{x}_0 \\
            \overline{y}_0
        \end{bmatrix} + \begin{bmatrix}
            \frac{1}{\lambda_1^*} \Sigma_{x_0} (y - x) \\
            \frac{1}{\lambda_2^*} \Sigma_{y_0} (x - y)
        \end{bmatrix}.
    \end{aligned}
\end{equation}
Now, applying the value of the primal solution in the second set of equations of~\eqref{eq:KKT-conditions}, we obtain

\begin{equation}
    \label{eq:latter-KKT-condition}
    \begin{aligned}
       &\begin{cases}
           \lambda_1^* \left[ \left\| s_1^* - \overline{x}_0 \right\|^{2}_{\Sigma_{x_0}^{-1}} - 1\right] = 0\\
           \lambda_2^* \left[ \left\| s_2^* - \overline{y}_0 \right\|^{2}_{\Sigma_{y_0}^{-1}}  - 1\right] = 0
       \end{cases} \implies
       &\begin{cases}
           {\lambda_1^*} = \left\| y - x \right\|_{\Sigma_{x_0}} \\
           {\lambda_2^*} = \left\| x - y \right\|_{\Sigma_{y_0}}
       \end{cases},
    \end{aligned}
\end{equation}
since, by the KKT conditions, $\lambda_{i}^{*} \geq 0$, for $i = 1, 2$. Substituting the values of $\lambda_{1}^{*}$ and $\lambda_{2}^{*}$ in~\eqref{eq:first-KKT-condition}, we can compute the primal solution $s^{*}$ analytically as

\begin{equation}
    \begin{aligned}
        s_1^* &= \overline{x}_0 + \frac{1}{\left\| y - x \right\|_{\Sigma_{x_0}}} \Sigma_{x_0} (y - x) \\
        s_2^* &= \overline{y}_0 + \frac{1}{\left\| x - y \right\|_{\Sigma_{y_0}}} \Sigma_{y_0} (x - y)
    \end{aligned}
\end{equation}
where each component, $s_1^*$ and $s_2^*$, of the primal solution is given by the center of the ellipsoid plus its corresponding normalized steepest descent direction.
\end{proof}
\noindent
With this, we conclude that the first step of the Frank-Wolfe Algorithm can be calculated analytically. Now, we must analyze the choice of stepsize $\alpha$ through line search, considering our objective function $f(x, y) = \| x - y \|^2$. We intend to find $\alpha$ such that

\begin{equation}
    \label{eq:alpha-condition}
    \begin{aligned}
        \alpha = \underset{\alpha \in [0,1]}{\arg\min} \ f\left(x^k + \alpha (s_{1}^{k} - x^k), y^k + \alpha (s_{2}^{k} - y^k)\right) .
    \end{aligned}
\end{equation}
\textbf{Proposition 3.}
The optimal step size through line search can be analytically computed as

\begin{equation}
\label{eq:alpha-minimum}
\begin{aligned}
    \alpha^{*} = \mbox{Proj}_{[0,1]}\left( {u^{k}}^T \frac{r^k}{\left\| r^k - p^k \right\|_{2}} \right) \ .
\end{aligned}
\end{equation}
where $r^{k} = x^{k} - y^{k}$ and $p^k = s_1^{k} - s_2^{k}$ are the range vector between the current estimations and its linearized approximation, respectively, the unit vector $u^{k} = \frac{r^k - p^k}{\left\| r^k - p^k \right\|_{2}}$ represents the mismatch between the range vector and its linearized approximation and $\mbox{Proj}_{C}(\cdot)$ is the projection operator onto set $C$.

\begin{proof}
Analyzing the update step at the $k$-th iteration for our problem and dropping $k$ for simplicity, we see that the objective function of our problem becomes

\begin{equation}
    \label{eq:alpha-computation}
    \begin{aligned}
        f(x + \alpha (s_1 - x), y + \alpha (s_2 - y)) &= \left\| x + \alpha (s_1 - x) - y - \alpha (s_2 - y)\right\|_{2}^2\\
        &= \left\| -\alpha \left[ (x - y) - (s_1 - s_2) \right] + (x - y)\right\|_{2}^2\\
        &= \alpha^2 \left\| (x - y) - (s_1 - s_2) \right\|_{2}^2 - 2 \alpha \left[ (x - y) - (s_1 - s_2) \right]^T (x - y) + \left\| x - y \right\|_{2}^2 \\
    \end{aligned}
\end{equation}
Since $\left\| (x - y) - (s_1 - s_2) \right\|_{2}^2$ is non-negative, then this function is convex in terms of $\alpha$. The minimum of the function occurs when the first derivative with respect to $\alpha$ is equal to zero, therefore

\begin{equation}
    \begin{aligned}
        \alpha^{*} = \frac{\left[ (x - y) - (s_1 - s_2) \right]^T (x - y)}{\left\| (x - y) - (s_1 - s_2) \right\|_{2}^2} = u^T \frac{r}{\left\| r - p \right\|_{2}} ,
    \end{aligned}
\end{equation}
where $r = x - y$, $p = s_1 - s_2$ and  $u^T = \frac{r - p}{\left\| r - p \right\|_{2}}$.

\textbf{Theorem 1 (Weierstrass Theorem)}
Let $a, b \in \mathbb{R}$ such that $a \leq b$ and let $\Psi$ be a continuous function in $[a,b]$. Then, exists $x_m, x_M \in [a,b]$, such that

\begin{equation}
    \label{eq:weierstrass}
    \begin{aligned}
        \forall x \in [a,b] : \Psi(x_m) \leq \Psi(x) \leq \Psi(x_M) .
    \end{aligned}
\end{equation}
\noindent
By Weierstrass' theorem, we have that this function of $\alpha$ has a minimum in the range $[0,1]$. Since the function is quadratic convex, there are three cases:

\begin{itemize}
    \item If the $\alpha$ calculated through~\eqref{eq:alpha-minimum}, is in the range, then $\alpha$ is already the minimum;
    \item If the calculated $\alpha$ is less than zero, then the minimum in the range $[0,1]$ occurs at $0$;
    \item If the calculated $\alpha$ is greater than $1$, then the minimum in the range $[0,1]$ occurs at $1$.
\end{itemize}
\end{proof}
\noindent
We conclude that both intermediate steps of the Frank-Wolfe algorithm can be computed analytically, thus presenting a fast alternative to computing the satellite safe margin. A summary of the full implementation of the Frank-Wolfe algorithm for our specific problem is presented in Algorithm~\ref{alg:margin-line-search-frank-wolfe}.

\begin{algorithm}[ht!]
\caption{Satellite Safe Margin with Frank-Wolfe Algorithm}
\label{alg:margin-line-search-frank-wolfe}
\begin{algorithmic}
\Require{$x, y$ such that $(x - \overline{x}_0)^T \Sigma_{x_0}^{-1} (x - \overline{x}_0) \leq 1$ and $(y - \overline{y}_0)^T \Sigma_{y_0}^{-1} (y - \overline{y}_0) \leq 1$}
\State $x^0 \gets z$
\State $y^0 \gets y$

\For{$k = 0,...,n-1$}
    \State $s_1^k \gets \overline{x}_0 + \frac{\Sigma_{x_0} (y^k - x^k)}{\left\| y^k - x^k \right\|_{\Sigma_{x_0}}}$ 
    \State $s_2^k \gets \overline{y}_0 + \frac{\Sigma_{y_0} (x^k - y^k)}{\left\| x^k - y^k \right\|_{\Sigma_{y_0}}}$
    \State $u^{k} \gets \frac{(x^{k} - y^{k}) - (s_1^{k} - s_2^{k})}{\left\| (x^{k} - y^{k}) - (s_1^{k} - s_2^{k}) \right\|_{2}^2}$
    \State $\alpha \gets \mbox{Proj}_{[0,1]}\left( {u^{k}}^T (x^{k} - y^{k}) \right)$
    
    \State $x^{k+1} \gets x^k + \alpha (s_1^k - x^k)$
    \State $y^{k+1} \gets y^k + \alpha (s_2^k - y^k)$
\EndFor \\

\Return $x^n, y^n$
\end{algorithmic}
\end{algorithm}

\section{Iterative solution using Fast Iterative Shrinkage-Thresholding Algorithm}
\label{sec:fista}

In cases of conjunctions between satellites belonging to different operators, they may not be willing to share the most accurate information (for example, the covariance matrix) regarding the position of objects. Therefore, it is important to allow the calculation of the \emph{margin} in line with the interests of operators in protecting their exclusive information. So, we present a way to compute it in a distributed way, where the covariance matrices of the objects are not shared, but only the points of possible position at each iteration. In this setting, we resort to the Fast Iterative Shrinkage-Thresholding Algorithm (FISTA).

\subsection{Fast Iterative Shrinkage-Thresholding Algorithm}

\noindent Consider the general problem formulation

\begin{equation}
    \label{eq:fista-general-problem-formulation}
    \begin{aligned}
        &\underset{x \in \mathbb{R}^n}{\mbox{minimize}} \quad \Psi(x) + \Omega(x)
    \end{aligned}
\end{equation}
with the following assumptions:

\begin{itemize}
    \item[$\bullet$] $\Omega: \mathbb{R}^n \rightarrow \mathbb{R}$ is convex (possibly non-smooth);
    \item[$\bullet$] $\Psi: \mathbb{R}^n \rightarrow \mathbb{R}$ is proper and closed, dom($\Psi$) is convex, dom($\Omega$) $\subseteq$ int(dom($\Psi$)), and $\Psi$ is $L_{\Psi}$-smooth over int(dom($\Psi$));
    \item[$\bullet$] The optimal set of problem~\eqref{eq:fista-general-problem-formulation} is nonempty.
\end{itemize}
A simple way to minimize the function $\Psi$, and with low computational cost, is to use the gradient algorithm whose iteration step is defined as

\begin{equation}
    \label{eq:iteration-step}
    \begin{aligned}
        x^{k+1} = x^k - \frac{1}{L_k} \nabla \Psi(x^k),
    \end{aligned}
\end{equation}
where $x^k$ and $L_k$ denotes the values of $x$ and $L$ at iteration $k$, with $\frac{1}{L_k} > 0$ defining the step-size. Since the $\Omega$ function can be non-smooth, the proximal operator is normally used, which can be defined as

\begin{equation}
    \label{eq:proximal-operator}
    \begin{aligned}
        \mbox{prox}_{\Omega}(x) = \underset{u \in \mathbb{R}^n}{\arg\min} \ \left\{\Omega(u) + \frac{1}{2} \|u - x\|^2\right\}.
    \end{aligned}
\end{equation}
The idea of the proximal operator is, given a point $x$, we aim to find the closest point to $x$ that minimizes the function $\Omega$. When $\Omega$ is properly closed and convex, $\mbox{prox}_{\Omega} (\cdot)$ exists and is unique \cite{beck2017first}. 

The proximal gradient method solves Problem~\eqref{eq:fista-general-problem-formulation} by combining the gradient step and the proximal operator by defining the iterative step

\begin{equation}
    \label{eq:iteration-step-prox-operator}
    \begin{aligned}
        x^{k+1} =\mbox{prox}_{\frac{1}{L_k} \Omega}\left(x^k - \frac{1}{L_k} \nabla \Psi(x^k)\right).
    \end{aligned}
\end{equation}
Thus, the Proximal Gradient method is defined as follows: pick $x^0 \in \mbox{int(dom}(\Psi))$ and iteratively apply the update step~\eqref{eq:iteration-step-prox-operator}. Analyzing some particular cases, when $\Omega \equiv 0$, then the update step reduces to the gradient step as described in~\eqref{eq:iteration-step}. When $\Omega = \delta_C$, where $\delta_C$ is the indicator function of the set $C$, written as

\begin{equation}
    \label{eq:general-indicator-function}
    \begin{aligned}
        \delta_C(x) = \begin{cases}
        0 & \quad \mbox{if } x \in C \\
        +\infty & \quad \mbox{otherwise}
        \end{cases} \ ,
    \end{aligned}
\end{equation}
such that $C$ is a nonempty closed and convex set, then the proximal operator becomes the orthogonal projection operator, $P_C$, which projects a point to the set C \cite{beck2017first}.

However, the proximal gradient method achieves a rate of convergence of $O\left(\frac{1}{k}\right)$. \citeauthor{nesterov1983method} presented a method based on a multistep scheme with a rate of convergence of $O\left(\frac{1}{k^2}\right)$ for this class of problems and proved that this rate of convergence is optimal and cannot be improved \cite{nesterov1983method}. \citeauthor{beck2009fast} present the Fast Iterative Shrinkage-Thresholding Algorithm (FISTA) \cite{beck2009fast}, which has a variant of the multistep scheme that achieves the same theoretical optimal convergence rate. A description of the FISTA algorithm can be found in Algorithm~\ref{alg:fista}.

\begin{algorithm}[h!]
\caption{Fast Iterative Shrinkage-Thresholding Algorithm (FISTA)}
\label{alg:fista}
\begin{algorithmic}
\Require{$p \in \mbox{int(dom}(\Psi))$, $\Psi$ and $\Omega$ with assumptions made for Problem~\eqref{eq:fista-general-problem-formulation}}
\State $x^0 \gets p$
\State $z^0 \gets x^0$
\State $t^0 \gets 1$

\For{$k = 0,...,n$}
    \State pick $L_k$ as Lipschitz constant of the gradient of the loss ($L_{\Psi}$) or value found by backtracking
    \State $x^{k+1} \gets \mbox{prox}_{\frac{1}{L_k} \Omega}\left(z^k - \frac{1}{L_k} \nabla \Psi(z^k) \right)$
    \State $t^{k+1} \gets \frac{1 + \sqrt{1 + 4 (t^k)^2}}{2}$
    \State $z^{k+1} \gets x^{k+1} + \left(\frac{t^k - 1}{t^{k+1}}\right) (x^{k+1} - x^k)$
\EndFor
\end{algorithmic}
\end{algorithm}
The main difference is that the proximal operator is not applied on the previous point $x^{k-1}$. Instead, it is applied on $z^k$ which is a specific linear combination of the previous two iterations, $x^{k-1}$ and $x^{k-2}$ \cite{beck2009fast}. This multistep scheme allows us to obtain the optimal convergence rate of $O\left(\frac{1}{k^2}\right)$. The choice of stepsize $\frac{1}{L_k}$ can be done in two ways:

\paragraph*{Constant stepsize:}

Remembering that the function $\Psi$ is $L_{\Psi}$-smooth, we can choose $L_k$ as the Lipschitz constant, $L_{\Psi}$, of the gradient of the function $\Psi$, so $L_k = L_{\Psi}$ for all $k$.

\paragraph*{Backtracking procedure:}

Choose $s > 0$ and $\eta > 1$. Define $L_{-1} = s$. At iteration $k$ $(k \geq 0)$, we choose $L_k$ as follows: first, set $L_k = L_{k-1}$. While

\begin{equation}
    \begin{aligned}
        \Psi\left( T_{L_k}( z^k ) \right) > \Psi(z^k) + \nabla \Psi(z^k)^T \left( T_{L_k}(z^k) - z^k \right) + \frac{L_k}{2} \left\| T_{L_k}(z^k) - z^k \right\|^2,
    \end{aligned}
\end{equation}
we set $L_k = \eta L_k$, such that

\begin{equation}
    \begin{aligned}
        T_{L_k}(x) = \mbox{prox}_{\frac{1}{L_k} \Omega}(x - \frac{1}{L_k} \nabla \Psi(x)).
    \end{aligned}
\end{equation}

\subsection{Optimal and stable margin computation without shared covariances}

Let us return to our original problem expressed in~\eqref{eq:margin-optimization-problem}. Using the indicator function, we can transform it into an unconstrained minimization problem as

\begin{equation}
    \label{eq:unconstrained-margin-optimization-problem}
    \begin{aligned}
        &\underset{x,y}{\mbox{minimize}} \quad \| x - y \|^2 +  \delta_{\xi_x}(x) + \delta_{\xi_y}(y),
    \end{aligned}
\end{equation}
where $\xi_x$ and $\xi_y$ are the error ellipses of the position of the objects and $\delta_{\xi_x}$ and $\delta_{\xi_y}$ are the corresponding indicator functions of the error ellipsoids. Visualizing our objective function as $f(x, y) = \| x - y \|^2$ and $g(x, y) = \delta_{\xi_x}(x) + \delta_{\xi_y}(y)$, we are in the same context as the general problem for the application of the FISTA method. The update step in~\eqref{eq:iteration-step-prox-operator} for this specific problem becomes

\begin{equation}
    \label{eq:ellipse-iteration-step-prox-operator}
    \begin{aligned}
        \begin{bmatrix}
            x^{k+1} \\
            y^{k+1}
        \end{bmatrix} &= \mbox{prox}_{\frac{1}{L_k} g}\left(\begin{bmatrix}
            x^k \\
            y^k
        \end{bmatrix} - \frac{1}{L_k} \nabla f\left( x^k, y^k \right)\right) \\
        &= \mbox{prox}_{\frac{1}{L_k} g}\left(\begin{bmatrix}
            x^k \\
            y^k
        \end{bmatrix} - \frac{1}{L_k} \begin{bmatrix}
            \frac{\partial}{\partial x}\left( \left\| x^k - y^k \right\|^2 \right) \\
            \frac{\partial}{\partial y}\left( \left\| x^k - y^k \right\|^2 \right)
        \end{bmatrix}\right) \\
        &= \mbox{prox}_{\frac{1}{L_k} g}\left(\begin{bmatrix}
            x^k \\
            y^k
        \end{bmatrix} - \frac{1}{L_k} \begin{bmatrix} 2 \left( x^k - y^k \right) \\
        2 \left( y^k - x^k \right)
        \end{bmatrix}\right).
    \end{aligned}
\end{equation}
Since $g(x,y) = \delta_{\xi_x}(x) + \delta_{\xi_y}(y)$, then the proximal operator becomes the orthogonal projection operator, which translates to projecting the point $x$ onto the ellipse $\xi_x$ and the point $y$ onto the ellipse $\xi_y$, at each iteration $k$.

However, we see that this can be uncoupled. Suppose at each iteration $k$, the chaser communicates its point extrapolation point $z^k$, which is a linear combination of the two previous iterates, and the target also communicates its extrapolation point. In that case, each agent can calculate the iteration step and project the point to its own ellipse. So each agent can implement its version of the FISTA method.

In the algorithm description, $\xi_i$ represents the error ellipse of each agent $i$. The operator $P_{\xi_i}$ is the orthogonal projection operator that projects a point onto the ellipse $\xi_i$. Later we will demonstrate how to compute this projection of a point onto the ellipse. For the implementation of the algorithm, we decided to choose $L_k = L_f$ since this constant is known for this problem.

\textbf{Fact.}
Let $\Psi: \mathbb{R}^n \rightarrow \mathbb{R}$ be twice differentiable and if there exists $L_{\Psi} < \infty$ such that its Hessian matrix has a bounded spectral norm:

\begin{equation}
    \begin{aligned}
    \opnorm{\nabla^2 \Psi(x)} \leq L_{\Psi}, \quad \forall x \in \mathbb{R}^{n},
    \end{aligned}
\end{equation}
such that $\opnorm{\cdot}$ denotes the spectral norm, then $\Psi$ has a Lipschitz continuous gradient with Lipschitz constant $L_{\Psi}$.
\noindent Computing the Hessian Matrix of our objective function $f$, we obtain

\begin{equation}
    \label{eq:f-hessian}
    \begin{aligned}
    \nabla^2 f(x,y) = \begin{bmatrix}
        2 I & -2 I \\
        -2 I & 2 I
    \end{bmatrix}
    \end{aligned}
\end{equation}
such that $I$ denotes the identity matrix. The spectral norm is the greatest singular value, which for the case of this Hessian matrix is 4, so we have $L_f = 4$. A complete description of the algorithm applied by each agent can be found in Algorithm \ref{alg:cap}. The optimal solution of each agent will be stored in the variable $x_k \in \mathbb{R}^{3}$.

\begin{algorithm}
\caption{FISTA at agent $i$, for $i \in \{1, 2\}$}\label{alg:cap}
\begin{algorithmic}
\Require{$n > 0$, $d \in \xi_i$}
\State $x^0 \gets d$
\State $p_i^0 \gets x^0$
\State $t^0 \gets 1$

\For{$k = 0,...,n-1$}
    \State Communicate $p_i^k$ and receive $p_j^k$ from agent $j$, such that $j \not\eq i$ and $j \in \{1, 2\}$
    \State $L_k \gets 4 \quad \mbox{(by constant stepsize)}$
    \State $x^{k+1} \gets P_{\xi_i}\left(p_i^k - \frac{2}{L_k} (p_i^k - p_j^k) \right)$
    \State $t^{k+1} \gets \frac{1 + \sqrt{1 + 4 (t^k)^2}}{2}$
    \State $p_i^{k+1} \gets x^{k+1} + \left(\frac{t^k - 1}{t^{k+1}}\right) (x^{k+1} - x^k)$
\EndFor \\
\Return $x^n$
\end{algorithmic}
\end{algorithm}

\subsection{Projection onto the ellipse}

\noindent Consider the constraint set

\begin{equation}
    \label{eq:constraint-set}
    \begin{aligned}
    C = \left\{ y \in \mathbb{R}^n : \sum_{i=1}^n w_i^2 (y_i - r_i)^2 \leq \epsilon^2 \right\}
    \end{aligned}
\end{equation}
This constraint set can be seen as an ellipse aligned with the axes. Let $x \not\in C$. The projection of $x$ onto $C$, $g = P_C(x)$, is on the boundary of C, so the boundary point $y$ must satisfy

\begin{equation}
    \label{eq:boundary-constraint}
    \begin{aligned}
    \sum_{i=1}^n w_i^2 (y_i - r_i)^2 = \epsilon^2
    \end{aligned}
\end{equation}
Thus, we see that the projection $g$ intends to minimize $\|x -y\|$ among all points $y$ that satisfy the interior condition. Calculating the Lagrangian

\begin{equation}
    \label{eq:lagrangian-constraint}
    \begin{aligned}
    L(y, \lambda) = \|x-y\|^2 + \lambda \left( \sum_{i=1}^n w_i^2 (y_i - r_i)^2 - \epsilon^2 \right),
    \end{aligned}
\end{equation}
and computing the partial derivative with respect to a particular $y_i$ and setting it to zero, we obtain

\begin{equation}
    \label{eq:boundary-point}
    \begin{aligned}
    y_i = \frac{x_i + \lambda w_i^2 r_i}{1 + \lambda w_i^2}.
    \end{aligned}
\end{equation}
Therefore, the projection $g \coloneqq (g_1,...,g_n)$ is given by

\begin{equation}
    \label{eq:projection-point}
    \begin{aligned}
    g_i = \frac{x_i + \lambda w_i^2 r_i}{1 + \lambda w_i^2}, \quad \mbox{for } i = 1,...,n,
    \end{aligned}
\end{equation}
such that $\lambda$, invoking the condition~\eqref{eq:boundary-constraint}, satisfies

\begin{equation}
    \label{eq:lambda-constraint}
    \begin{aligned}
    \sum_{i=1}^n \frac{w_i^2 (x_i - r_i)^2}{(1 + \lambda w_i^2)^2} = \epsilon^2
    \end{aligned}
\end{equation}
It is shown in \cite{stark1998vector} that the $\lambda$ that gives the correct projection is the only positive root of the equation~\eqref{eq:lambda-constraint}. The following theorem, also described in \cite{stark1998vector}, summarizes that the application of the Newton method guarantees convergence to the only positive root of the equation~\eqref{eq:lambda-constraint}.

\textbf{Theorem 2.}
Let function $\psi : \mathbb{R} \rightarrow \mathbb{R}$ be the residual of Eq.~\eqref{eq:lambda-constraint},

\begin{equation}
    \label{eq:lambda-constraint-function}
    \begin{aligned}
    \psi(\lambda) = \left( \sum_{i=1}^n \frac{w_i^2 (x_i - r_i)^2}{(1 + \lambda w_i^2)^2} \right) - \epsilon^2
    \end{aligned}
\end{equation}
such that $\epsilon^2 < \sum_{i=1}^n w_i^2 (x_i - r_i)$. Then, with $\lambda_0 = 0$ the iterates generated by Newton's method

\begin{equation}
    \label{eq:newton-iteration}
    \begin{aligned}
    \lambda_{k+1} = \lambda_k - \frac{\psi(\lambda)}{\psi^{'} (\lambda)}, \quad \mbox{for } k = 1,2,...,
    \end{aligned}
\end{equation}
will converge increasingly to $\lambda_+$, the unique positive root of $\psi(\lambda) = 0$.

The fact that Newton's method converges to the correct value ensures a fast algorithm for projecting a point on the ellipse since Newton's method has quadratic convergence. Proofs about the convergence of Newton's method to the correct value of $\lambda$ and about the guarantee of correct projection can be found in \cite{stark1998vector}.

\section{Numerical Experiments and Discussion}
\label{sec:experiments}

In this section, we present the results of the numerical experiments carried out. We present the results of the accuracy and processing time of the two approaches described above, using the Frank-Wolfe Algorithm and Fast Iterative Shrinkage-Thresholding Algorithm. We compare these two approaches with the results obtained with the CVXPY, an open source Python-embedded modeling language for convex optimization problems \cite{diamond2016cvxpy,agrawal2018rewriting}, which we use as ground truth, and also with the results obtained through the analytical approach presented by \citeauthor{rimon1997obstacle} \cite{rimon1997obstacle}, which was also described previously.

\subsection{Dataset}

For the numerical experiments, we used the real Conjunction Data Message (CDM) dataset used in ESA's Collision Avoidance Challenge \citep{uriot2021spacecraft}. More details about the dataset can be found on the challenge's page \footnote{\url{https://kelvins.esa.int/collision-avoidance-challenge/data/}}.

The highest degree of uncertainty is associated with the transverse (or along-track) component. This happens due to the impact of non-conservative and perturbative forces on the motion of the object and also due to the high velocities of objects in Low-Earth Orbit (LEO), which lead to a greater uncertainty when measuring the object's position with sensors on the ground \citep{vallado2001fundamentals}. To remove some outliers and unrealistic cases like those where the transverse component standard deviation is greater than the Earth's radius, we decided to remove samples beyond the $95$-th percentile of the chaser's transverse component standard deviation, given that the chaser reveals a greater uncertainty associated with the position. We then select 20,000 samples from this dataset to test our solutions to compute the \emph{margin}.

\paragraph*{Ground Truth with CVXPY:}

To guarantee a ground truth for our problem we use CVXPY, a general purpose convex solver \cite{diamond2016cvxpy,agrawal2018rewriting}. With this general solver, we obtain the optimal solution and use these results and their processing time as a basis to compare with the proposed approaches. The average CVXPY processing time for each sample is about $158$ ms.

\subsection{Rimon-Boyd Benchmark}

As presented earlier, Rimon and Boyd present an analytical method for calculating the distance between ellipses. This method could be used to calculate the \emph{margin} in a centralized way. However, this method gives wrong results in some cases. As explained earlier, non-normal matrices are extremely sensitive to numerical and/or round-off errors in the computation of their eigenvalues and these may not be accurate. An inaccurate eigenvalue implies a wrong result in the estimation of the distance between ellipses since the result depends exclusively on the computation of the smallest eigenvalue of the matrices expressed in~\eqref{eq:min-eigenvalue-1} and~\eqref{eq:min-eigenvalue-2}. 

\begin{figure}[ht!]
    \centering
    \includegraphics[scale=0.8]{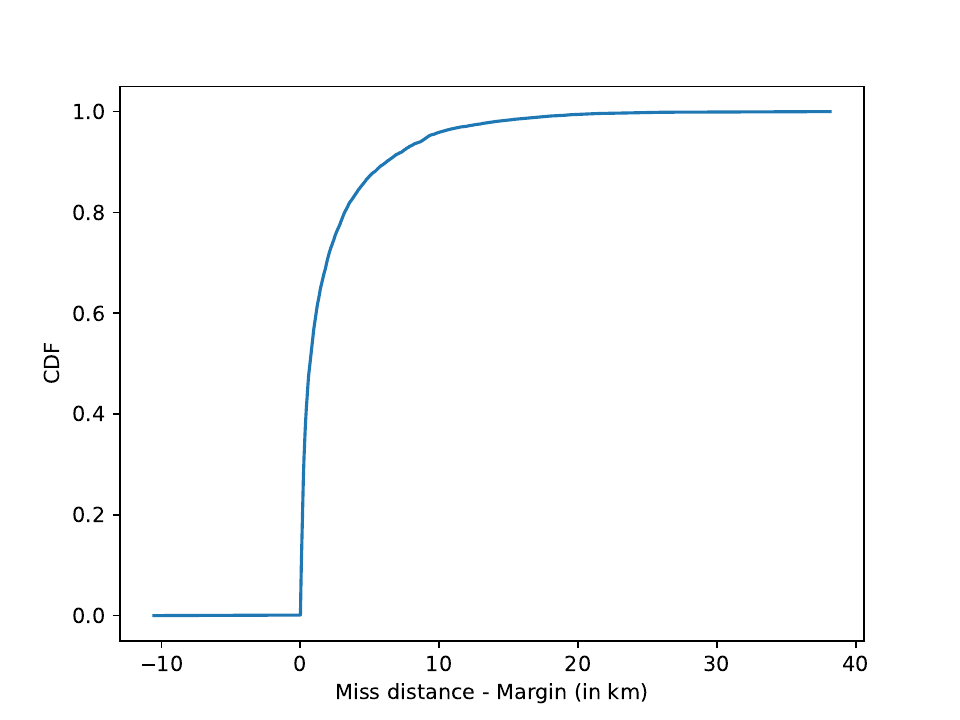}
    \caption[Cumulative Distribution Function of the difference between the miss distance and the margin computed with the Rimon and Boyd method.]{Cumulative Distribution Function of the difference between the miss distance and the margin computed with the Rimon and Boyd method.} 
    \label{fig:rimon-boyd-cdf}
\end{figure}
In Fig.~\ref{fig:rimon-boyd-cdf}, we have the cumulative distribution function (CDF) of the difference between the miss distance and the margin computed with the Rimon and Boyd method. For each value in the $x$-axis, we have the information of the percentage of cases where the difference between the miss distance and the computed margin is less or equal to $x$. In this case, we see that there are a few cases where this difference is negative, therefore the computed margin is greater than the miss distance, which is conceptually impossible.

To assess the error in the precision of the calculated value, we compute the difference between the estimated value and the ground truth as

\begin{equation}
    \label{eq:relative-error-metric}
    \begin{aligned}
    \varepsilon = \hat{x} - x^*
    \end{aligned}
\end{equation}
such that $\hat{x}$ is the value of \emph{margin} predicted by the approach presented and $x^*$ is the optimal value of \emph{margin} computed by CVXPY. We use the estimates of CVXPY as ground truth to assess the accuracy of each of the solutions presented. 

\begin{figure}[ht!]
    \centering
    \includegraphics[scale=0.8]{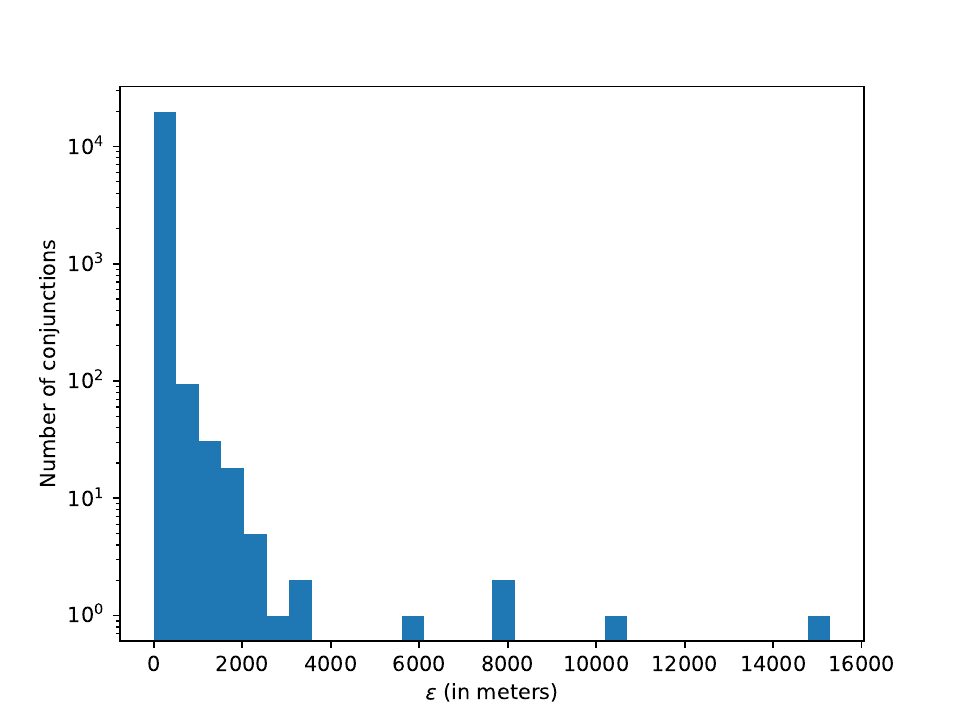}
    \caption[Relative error of the margin value computed with the Rimon-Boyd Approach.]{Relative error of the margin value computed with the Rimon-Boyd Approach.} 
    \label{fig:margin-boyd-rel-error}
\end{figure}

In Fig.~\ref{fig:margin-boyd-rel-error}, it is possible to observe that the estimation error $\varepsilon$ is high reaching an error of up to $16$ km. Despite the fast processing time of this method, of about $0.8$ ms per sample, this method presents high precision errors which are critical in the evaluation of conjunctions and protection of assets in orbit such as satellites. Therefore, this method does not present a viable solution for operators to compute this new robust solution.

\subsection{Frank-Wolfe and Fast Iterative Shrinkage-Thresholding Algorithms}

\begin{figure}[ht!]
    \centering
    \includegraphics[scale=0.8]{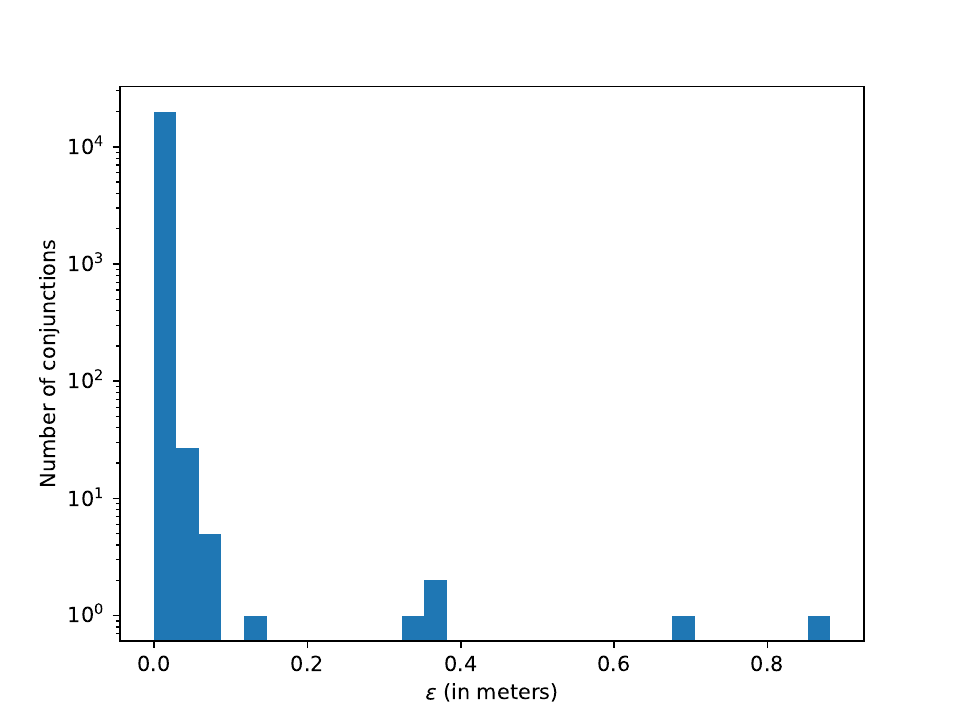}
    \caption[Relative error of the margin value computed with the Frank-Wolfe with Step-Size Line Search Approach.]{Relative error of the margin value computed with the Frank-Wolfe with Step-Size Line Search Approach.} 
    \label{fig:margin-fw-rel-error}
\end{figure}

The approach with the Frank-Wolfe Algorithm with Step-Size Line Search presents itself as a good alternative to calculate the \emph{margin} in a centralized way. For each sample, the algorithm runs for a number $k$ of iterations until the improvement is no greater than 1 mm, i.e., $\|x^k - x^{k-1}\| \leq 10^{-3}$. This threshold can be adapted by operators in their systems, keeping in mind that there is a trade-off between accuracy and processing time. As can be seen in Fig.~\ref{fig:margin-fw-rel-error}, this approach with the described stop-criterion gives more accurate results reaching a maximum estimation error of $0.8$ meters, with the majority of the cases having an error very close to zero. These results demonstrate the accuracy and robustness of the Frank-Wolfe Algorithm in computing the margin in different cases, in contrast to the method of Rimon and Boyd.

Regarding the processing time, the algorithm obtains an average processing time of $4$ ms. Despite this approach being approximately four times slower than the analytical approach proposed by Rimon and Boyd, it has the advantage of avoiding numerical errors and, consequently, obtaining accurate results. When compared to the processing time of the general solver CVXPY, the approach of the Frank-Wolfe Algorithm with Step-Size Line Search is almost 40 times faster.

Secondly, we present the numerical experiments and the results obtained with the approach using the Fast Iterative Shrinkage-Thresholding Algorithm, which is presented as an alternative to compute the \emph{margin} in a distributed way, without the agents needing to communicate their covariances, but only the prediction at the current iteration. Similar to the Frank-Wolfe Algorithm, we use the same stopping criterion.

\begin{figure}[ht!]
    \centering
    \includegraphics[scale=0.8]{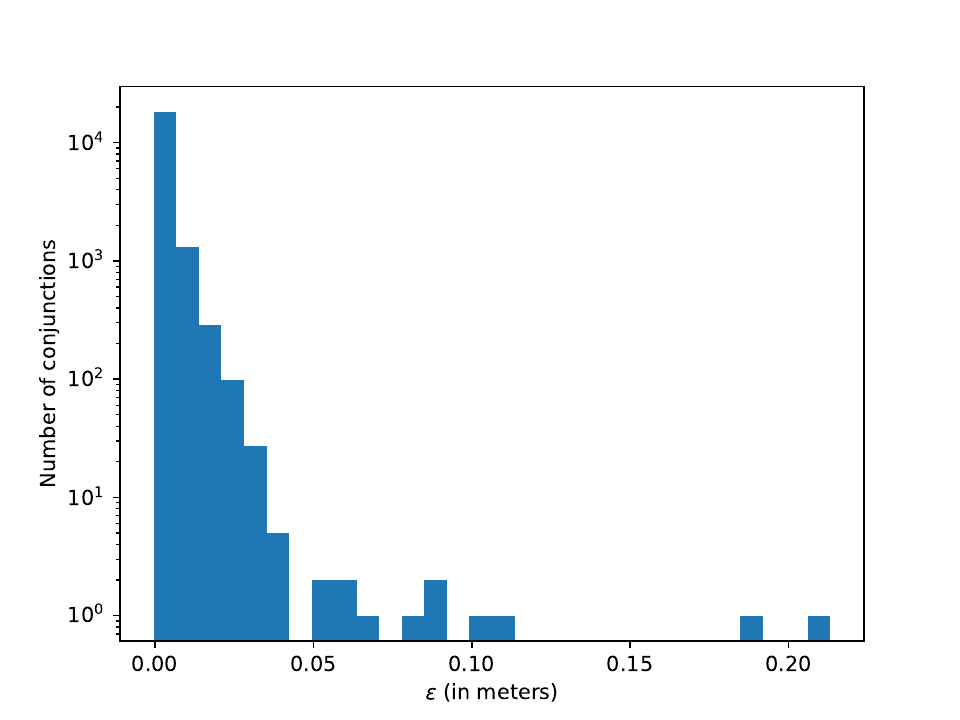}
    \caption[Relative error of the margin value computed with FISTA.]{Relative error of the margin value computed with the FISTA approach.} 
    \label{fig:margin-fista-rel-error}
\end{figure}

In Fig.~\ref{fig:margin-fista-rel-error}, we can see that this approach presents a maximum estimation error of 0.2 meters, with the majority of the cases having an error very close to zero. These results demonstrate the accuracy and robustness of the Fast Iterative Shrinkage-Thresholding Algorithm to compute the margin in different conjunctions, in contrast to the method of Rimon and Boyd. Regarding the processing time, the distributed approach presents an average processing time of $122$ ms, slightly faster than the general solver.

\subsection{Identification of cases of concern}

\begin{figure}[ht!]
    \centering
    \includegraphics[scale=0.8]{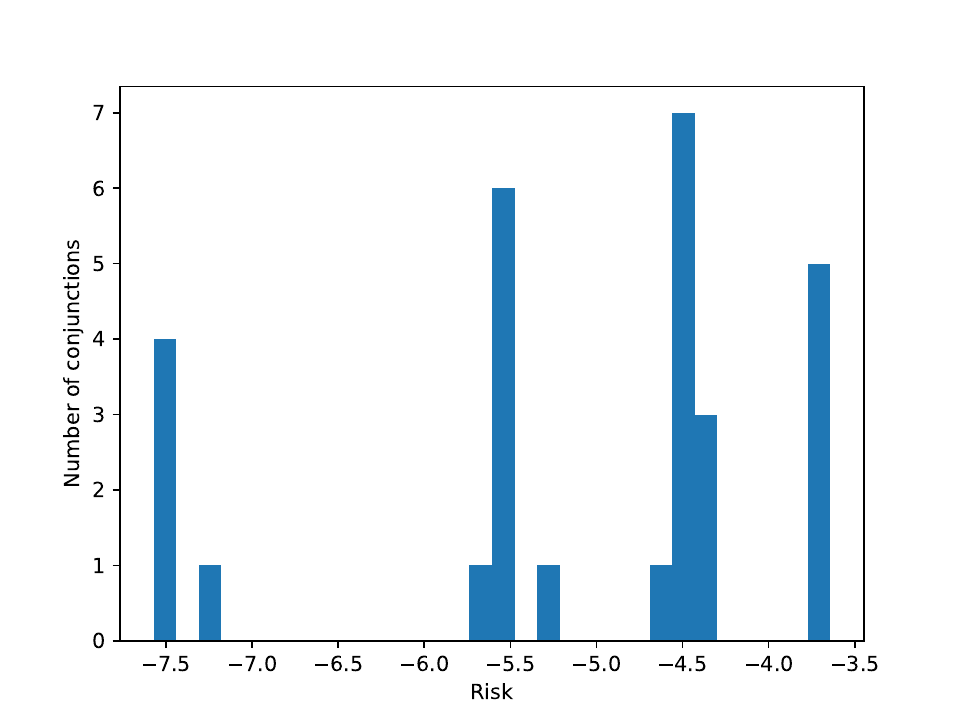}
    \caption[Histogram of the risk of the conjunctions identified by comparing when the margin is smaller than the combined radius of the objects.]{Histogram of the risk of the conjunctions identified by comparing when the margin is smaller than the combined radius of the objects.} 
    \label{fig:cases-of-concern-margin}
\end{figure}

We now demonstrate the applicability of the margin to identify cases of concern. As advocated by~\citeauthor{balch2019satellite}, it is important to take into account the physical size of the orbiting objects and compare if the minimum distance between the ellipsoids is smaller than the combined radius of the objects~\cite{balch2019satellite}. So, we analyze the cases where the computed margin is smaller than the sum of the radius of the objects. In this experiment, we consider 1-$\sigma$ uncertainty ellipsoids. In Fig.~\ref{fig:cases-of-concern-margin}, we can see a histogram of the cases that meet this condition and the corresponding risk of collision contained in the conjunction data message (CDM). The risk is defined as $\log_{10}\left( P_C \right)$, where $P_C$ denotes the probability of collision. We see that all the cases identified when the margin is smaller than the combined radius of the objects have a risk greater than $-7.5$, thus meeting a criterion to be considered cases of concern. It is true that exist cases with a risk greater than $-7$ that are not identified, nevertheless these preliminary results demonstrate that there are no false alarms. Even in cases that are not identified, the margin can be used to better understand the contours of the situation. We are aware that many Conjunction Data Messages (CDMs) do not contain information about certain parameters, such as the space debris span, and that is why standard values are often used. This information must be taken into account when using this tool for conjunction analysis, and the confidence level must be different in these cases.

\subsection{Discussion}

We showed two fast and stable ways of computing the \emph{margin}. However, one might consider how this distance between two ellipses representing the uncertainty associated with the objects' position differs from other distances such as the Mahalanobis distance~\cite{mahalanobis1936generalised}, or divergences between distributions, such as the Kullback-Leibler divergence~\cite{kullback1951information}.

The major argument in favor of this new pseudo-semi-metric is that it has a physical meaning easily comprehended by operators. As exposed before, assuming that the true positions of the objects are contained in their corresponding uncertainty ellipsoids, the margin guarantees that the smallest possible distance between them is not smaller than its value. We present this new tool, not as a replacement for any other tool for conjunction assessment, but to complement and aid operators when analyzing conjunctions. When the ellipses overlap, the margin returns zero, thus it is necessary to use other tools to understand the contours of the situation. Otherwise, this tool can help by presenting a level of confidence for the operators. They can choose how much probability is contained in the ellipses and obtain the smallest possible distance between the objects, analyzing if they consider it safe enough or if it is better to maneuver. For example, if for 3-$\sigma$ uncertainty ellipsoids we have a margin of zero, an operator could also compute for 2-$\sigma$ and 1-$\sigma$, and analyze more thoroughly when necessary.

One could compute the Mahalanobis distance as in~\cite{park2019statistical} for the Time of Closest Approach (TCA), where we consider a new random variable such that its mean value is the miss distance and its covariance is the sum of the objects' covariance matrices, we could compute the Mahalanobis distance the origin and the ellipsoid described by the new random variable. However, simplistically, the Mahalanobis distance returns the amount of standard deviations that the ellipsoid is to the origin. This concept is much harder to interpret than the Euclidean distance between the ellipsoids. Regarding Kullback-Leibler divergence, it follows the same argument about interpretability as it is difficult to rationalize what it means in terms of the risk of the conjunction. Also, the Kullback-Leibler divergence is not symmetric thus the order by which we consider the objects returns two different results, which further contributes to the difficulty of interpreting this statistical distance for conjunctions assessment.

\section{Conclusion}

Tools such as the probability of collision and the miss distance, key components in the entire process of monitoring and evaluating orbital encounters, do not take into account worst-case scenarios. The new tool presented for the analysis of conjunctions, the \emph{margin}, consists of the minimum possible distance between objects given the uncertainty associated with their position. We highlight that the operators can choose how much probability is contained in the uncertainty ellipsoids, i.e., the margin can be computed for 1-$\sigma$, 2-$\sigma$, 3-$\sigma$ ellipsoids, or even bigger. We do not present this new tool as a replacement for any other tool for conjunction assessment, but instead to complement the decision-making process and aid operators analyze conjunctions.

We present two ways to calculate this new tool for two different situations: when the best-known data from both objects can be centralized (e.g. debris-satellite conjunctions or conjunctions
between satellites from the same operator) and when the most precise covariances cannot be shared (conjunctions of satellites owned by different operators). The last
option allows for collaboration between satellite operators when more precise information about assets is not to be
shared directly while not neglecting accuracy. These two solutions make it possible to compute this new tool accurately and expeditiously. Running in a MacBook Air 13 2020, with an 8-core Apple M1 processor, 8-core GPU, 16GB of RAM and a non-parallelized implementation in Python 3.10, with the centralized solution, it is possible to process approximately 15,000 conjunctions per minute, and with the distributed solution, it is possible to process approximately 490 conjunctions per minute.

From some preliminary experiments, we analyze the applicability of the margin to identify cases of concern. By comparing when the margin is smaller than the combined radius of the objects, we see that all the cases identified have a risk greater than  $-7.5$, meeting a criterion to be considered cases of concern.

As future work, we consider it can be interesting to explore the applicability of this idea of the distance between ellipsoids to long-term conjunctions, where we no longer consider the instant of the Time of Closest Approach (TCA), but instead, we consider a time interval for the whole encounter. Another avenue of research we consider important is the study of the impact of the stopping criteria and the chosen threshold in the operators' operations. We consider important a more thorough study on how the choice of different stopping criteria can impact the accuracy and processing time, and the consequences for the operators.

\section{Acknowledgments}

This work is supported by NOVA LINCS (UIDB/04516/2020) with the financial support of FCT.IP. This research was carried out under Project ``Artificial Intelligence Fights Space Debris'' Nº C626449889-0046305 co-funded by Recovery and Resilience Plan and NextGeneration EU Funds, www.recuperarportugal.gov.pt.

\begin{center}
    \includegraphics[scale=0.12]{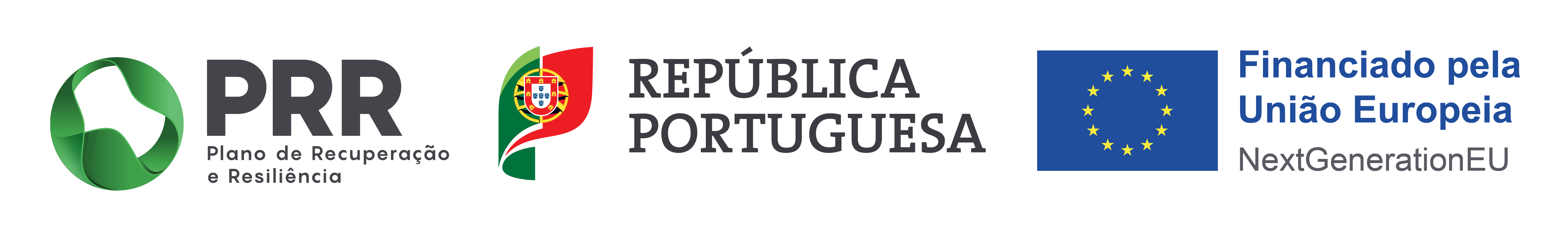}
\end{center}

\bibliography{sample}

\begin{thebibliography}{49}
\newcommand{\enquote}[1]{``#1''}
\providecommand{\natexlab}[1]{#1}
\providecommand{\url}[1]{\texttt{#1}}
\providecommand{\urlprefix}{URL }
\expandafter\ifx\csname urlstyle\endcsname\relax
  \providecommand{\doi}[1]{\discretionary{}{}{}https://doi.org/#1}\else
  \providecommand{\doi}[1]{\discretionary{}{}{}\urlstyle{rm}\url{https://doi.org/#1}}\fi

\bibitem[{Kessler and Cour-Palais(1978)}]{kessler1978collision}
Kessler, D.~J., and Cour-Palais, B.~G., \enquote{Collision frequency of
  artificial satellites: The creation of a debris belt,} \emph{Journal of
  Geophysical Research: Space Physics}, Vol.~83, No.~A6, 1978, pp. 2637--2646.
\newblock \doi{10.1029/JA083iA06p02637}.

\bibitem[{Pelton(2013)}]{pelton2013space}
Pelton, J.~N., \emph{Serious Threats from Outer Space}, Springer New York,
  2013, pp. 1--15.
\newblock \doi{10.1007/978-1-4614-6714-4_1}.

\bibitem[{Manfletti et~al.(2023)Manfletti, Guimarães, and
  Soares}]{manfletti2023ai}
Manfletti, C., Guimarães, M., and Soares, C., \enquote{AI for space traffic
  management,} \emph{Journal of Space Safety Engineering}, Vol.~10, No.~4,
  2023, pp. 495--504.
\newblock \doi{10.1016/j.jsse.2023.08.007}.

\bibitem[{Kleinig et~al.(2022)Kleinig, Smith, and Capon}]{kleinig2022collision}
Kleinig, T., Smith, B., and Capon, C., \enquote{Collision avoidance of
  satellites using ionospheric drag,} \emph{Acta Astronautica}, Vol. 198, 2022,
  pp. 45--55.
\newblock \doi{10.1016/j.actaastro.2022.03.017}.

\bibitem[{Bonnal et~al.(2020)Bonnal, McKnight, Phipps, Dupont, Missonnier,
  Lequette, Merle, and Rommelaere}]{bonnal2020just}
Bonnal, C., McKnight, D., Phipps, C., Dupont, C., Missonnier, S., Lequette, L.,
  Merle, M., and Rommelaere, S., \enquote{Just in time collision avoidance –
  A review,} \emph{Acta Astronautica}, Vol. 170, 2020, pp. 637--651.
\newblock \doi{10.1016/j.actaastro.2020.02.016}.

\bibitem[{Mishne and Edlerman(2017)}]{mishne2017collision}
Mishne, D., and Edlerman, E., \enquote{Collision-Avoidance Maneuver of
  Satellites Using Drag and Solar Radiation Pressure,} \emph{Journal of
  Guidance, Control, and Dynamics}, Vol.~40, No.~5, 2017, pp. 1191--1205.
\newblock \doi{10.2514/1.G002376}.

\bibitem[{Reiland et~al.(2021)Reiland, Rosengren, Malhotra, and
  Bombardelli}]{reiland2021assessing}
Reiland, N., Rosengren, A.~J., Malhotra, R., and Bombardelli, C.,
  \enquote{Assessing and minimizing collisions in satellite
  mega-constellations,} \emph{Advances in Space Research}, Vol.~67, No.~11,
  2021, pp. 3755--3774.
\newblock \doi{10.1016/j.asr.2021.01.010}.

\bibitem[{{Le May} et~al.(2018){Le May}, Gehly, Carter, and
  Flegel}]{le2018space}
{Le May}, S., Gehly, S., Carter, B., and Flegel, S., \enquote{Space debris
  collision probability analysis for proposed global broadband constellations,}
  \emph{Acta Astronautica}, Vol. 151, 2018, pp. 445--455.
\newblock \doi{10.1016/j.actaastro.2018.06.036}.

\bibitem[{Lucken and Giolito(2019)}]{lucken2019collision}
Lucken, R., and Giolito, D., \enquote{Collision risk prediction for
  constellation design,} \emph{Acta Astronautica}, Vol. 161, 2019, pp.
  492--501.
\newblock \doi{10.1016/j.actaastro.2019.04.003}.

\bibitem[{Balch et~al.(2019)Balch, Martin, and Ferson}]{balch2019satellite}
Balch, M.~S., Martin, R., and Ferson, S., \enquote{Satellite conjunction
  analysis and the false confidence theorem,} \emph{Proceedings of the Royal
  Society A: Mathematical, Physical and Engineering Sciences}, Vol. 475, No.
  2227, 2019, p. 20180565.
\newblock \doi{10.1098/rspa.2018.0565}.

\bibitem[{Poore et~al.(2016)Poore, Aristoff, Horwood, Armellin, Cerven, Cheng,
  Cox, Erwin, Frisbee, Hejduk, Jones, Di~Lizia, Scheeres, Vallado, and
  Weisman}]{poore2016covariance}
Poore, A.~B., Aristoff, J.~M., Horwood, J.~T., Armellin, R., Cerven, W.~T.,
  Cheng, Y., Cox, C.~M., Erwin, R.~S., Frisbee, J.~H., Hejduk, M.~D., Jones,
  B.~A., Di~Lizia, P., Scheeres, D.~J., Vallado, D.~A., and Weisman, R.~M.,
  \enquote{Covariance and uncertainty realism in space surveillance and
  tracking,} Tech. rep., Numerica Corporation Fort Collins United States, 2016.
\newblock \urlprefix\url{https://apps.dtic.mil/sti/citations/AD1020892}.

\bibitem[{Klinkrad et~al.(2006)Klinkrad, Alarc{\'o}n, and
  S{\'a}nchez}]{klinkrad2006space}
Klinkrad, H., Alarc{\'o}n, J., and S{\'a}nchez, N., \emph{Operational Collision
  Avoidance with Regard to Catalog Objects}, Springer Berlin Heidelberg, 2006,
  pp. 215--240.
\newblock \doi{10.1007/3-540-37674-7_8}.

\bibitem[{Klinkrad et~al.(2005)Klinkrad, Alarcon, and
  Sanchez}]{klinkrad2005collision}
Klinkrad, H., Alarcon, J., and Sanchez, N., \enquote{Collision avoidance for
  operational ESA satellites,} \emph{4th European Conference on Space Debris},
  Vol.~4, 2005.
\newblock
  \urlprefix\url{https://conference.sdo.esoc.esa.int/proceedings/sdc4/paper/49}.

\bibitem[{nas(2020)}]{nasahandbook}
\enquote{Spaceflight Safety Handbook for Satellite Operators: 18 SPCS Processes
  for On-Orbit Conjunction Assessment \& Collision Avoidance,} , 2020.
\newblock \urlprefix\url{www.space-track.org}.

\bibitem[{Merz et~al.(2021)Merz, Siminski, Virgili, Braun, Flegel, Flohrer,
  Funke, Horstmann, Lemmens, Letizia et~al.}]{merzesa}
Merz, K., Siminski, J., Virgili, B.~B., Braun, V., Flegel, S., Flohrer, T.,
  Funke, Q., Horstmann, A., Lemmens, S., Letizia, F., et~al.,
  \enquote{{ESA}’s Collision Avoidance Service: Current status and special
  cases,} \emph{Proceeding of the 8th European Conference on Space Debris
  (virtual)}, Vol.~8, 2021.
\newblock
  \urlprefix\url{https://conference.sdo.esoc.esa.int/proceedings/sdc8/paper/296}.

\bibitem[{Alfriend et~al.(1999)Alfriend, Akella, Frisbee, Foster, Lee, and
  Wilkins}]{alfriend1999probability}
Alfriend, K.~T., Akella, M.~R., Frisbee, J., Foster, J.~L., Lee, D.-J., and
  Wilkins, M., \enquote{Probability of collision error analysis,} \emph{Space
  Debris}, Vol.~1, 1999, pp. 21--35.
\newblock \doi{https://doi.org/10.1023/A:1010056509803}.

\bibitem[{Ferreira et~al.(2023)Ferreira, Soares, and
  Guimar{\~a}es}]{ferreira2023probability}
Ferreira, R., Soares, C., and Guimar{\~a}es, M., \enquote{Probability of
  Collision of satellites and space debris for short-term encounters:
  Rederivation and fast-to-compute upper and lower bounds,} \emph{74th
  International Astronautical Congress}, 2023.
\newblock
  \urlprefix\url{https://iafastro.directory/iac/archive/browse/IAC-23/A6/IP/76012/}.

\bibitem[{Alfano and Greer(2003)}]{alfano2003determining}
Alfano, S., and Greer, M.~L., \enquote{Determining If Two Solid Ellipsoids
  Intersect,} \emph{Journal of Guidance, Control, and Dynamics}, Vol.~26,
  No.~1, 2003, pp. 106--110.
\newblock \doi{10.2514/2.5020}.

\bibitem[{Tracy et~al.(2023)Tracy, Howell, and
  Manchester}]{tracy2023differentiable}
Tracy, K., Howell, T.~A., and Manchester, Z., \enquote{Differentiable Collision
  Detection for a Set of Convex Primitives,} \emph{2023 IEEE International
  Conference on Robotics and Automation (ICRA)}, 2023, pp. 3663--3670.
\newblock \doi{10.1109/ICRA48891.2023.10160716}.

\bibitem[{Rimon and Boyd(1997)}]{rimon1997obstacle}
Rimon, E., and Boyd, S.~P., \enquote{Obstacle collision detection using best
  ellipsoid fit,} \emph{Journal of Intelligent and Robotic Systems}, Vol.~18,
  No.~2, 1997, pp. 105--126.
\newblock \doi{10.1023/A:1007960531949}.

\bibitem[{Canny(1986)}]{canny1986collision}
Canny, J., \enquote{Collision Detection for Moving Polyhedra,} \emph{IEEE
  Transactions on Pattern Analysis and Machine Intelligence}, Vol. PAMI-8,
  No.~2, 1986, pp. 200--209.
\newblock \doi{10.1109/TPAMI.1986.4767773}.

\bibitem[{Kockara et~al.(2007)Kockara, Halic, Iqbal, Bayrak, and
  Rowe}]{kockara2007collision}
Kockara, S., Halic, T., Iqbal, K., Bayrak, C., and Rowe, R., \enquote{Collision
  detection: A survey,} \emph{2007 IEEE International Conference on Systems,
  Man and Cybernetics}, 2007, pp. 4046--4051.
\newblock \doi{10.1109/ICSMC.2007.4414258}.

\bibitem[{Gilbert et~al.(1988)Gilbert, Johnson, and Keerthi}]{gilbert1988fast}
Gilbert, E., Johnson, D., and Keerthi, S., \enquote{A fast procedure for
  computing the distance between complex objects in three-dimensional space,}
  \emph{IEEE Journal on Robotics and Automation}, Vol.~4, No.~2, 1988, pp.
  193--203.
\newblock \doi{10.1109/56.2083}.

\bibitem[{Polyak(1964)}]{polyak1964some}
Polyak, B., \enquote{Some methods of speeding up the convergence of iteration
  methods,} \emph{USSR Computational Mathematics and Mathematical Physics},
  Vol.~4, No.~5, 1964, pp. 1--17.
\newblock \doi{10.1016/0041-5553(64)90137-5}.

\bibitem[{Nesterov(1983)}]{nesterov1983method}
Nesterov, Y., \enquote{A method for solving the convex programming problem with
  convergence rate O(1/$k^2$),} \emph{Proceedings of the USSR Academy of
  Sciences}, Vol. 269, 1983, pp. 543--547.

\bibitem[{Montaut et~al.(2024)Montaut, Le~Lidec, Petrik, Sivic, and
  Carpentier}]{montaut2024gjk++}
Montaut, L., Le~Lidec, Q., Petrik, V., Sivic, J., and Carpentier, J.,
  \enquote{GJK++: Leveraging Acceleration Methods for Faster Collision
  Detection,} \emph{IEEE Transactions on Robotics}, Vol.~40, 2024, pp.
  2564--2581.
\newblock \doi{10.1109/TRO.2024.3386370}.

\bibitem[{Diamond and Boyd(2016)}]{diamond2016cvxpy}
Diamond, S., and Boyd, S., \enquote{CVXPY: A Python-Embedded Modeling Language
  for Convex Optimization,} \emph{Journal of Machine Learning Research},
  Vol.~17, No.~83, 2016, pp. 1--5.
\newblock \urlprefix\url{http://jmlr.org/papers/v17/15-408.html}.

\bibitem[{Akshay~Agrawal and Boyd(2018)}]{agrawal2018rewriting}
Akshay~Agrawal, S.~D., Robin~Verschueren, and Boyd, S., \enquote{A rewriting
  system for convex optimization problems,} \emph{Journal of Control and
  Decision}, Vol.~5, No.~1, 2018, pp. 42--60.
\newblock \doi{10.1080/23307706.2017.1397554}.

\bibitem[{Mahalanobis(1936)}]{mahalanobis1936generalised}
Mahalanobis, P., \enquote{On the generalised distance in statistics,}
  \emph{Proceedings of the National Institute of Science of India}, Vol.~12,
  1936, pp. 49--55.

\bibitem[{Kullback and Leibler(1951)}]{kullback1951information}
Kullback, S., and Leibler, R.~A., \enquote{{On Information and Sufficiency},}
  \emph{The Annals of Mathematical Statistics}, Vol.~22, No.~1, 1951, pp. 79 --
  86.
\newblock \doi{10.1214/aoms/1177729694}.

\bibitem[{Gilitschenski and Hanebeck(2012)}]{gilitschenski2012robust}
Gilitschenski, I., and Hanebeck, U.~D., \enquote{A robust computational test
  for overlap of two arbitrary-dimensional ellipsoids in fault-detection of
  Kalman filters,} \emph{2012 15th International Conference on Information
  Fusion}, 2012, pp. 396--401.

\bibitem[{Brent(1971)}]{brent1971algorithm}
Brent, R.~P., \enquote{An algorithm with guaranteed convergence for finding a
  zero of a function,} \emph{The Computer Journal}, Vol.~14, No.~4, 1971, pp.
  422--425.
\newblock \doi{10.1093/comjnl/14.4.422}.

\bibitem[{Virtanen et~al.(2020)Virtanen, Gommers, Oliphant, Haberland, Reddy,
  Cournapeau, Burovski, Peterson, Weckesser, Bright, {van der Walt}, Brett,
  Wilson, Millman, Mayorov, Nelson, Jones, Kern, Larson, Carey, Polat, Feng,
  Moore, {VanderPlas}, Laxalde, Perktold, Cimrman, Henriksen, Quintero, Harris,
  Archibald, Ribeiro, Pedregosa, {van Mulbregt}, and {SciPy 1.0
  Contributors}}]{2020SciPy-NMeth}
Virtanen, P., Gommers, R., Oliphant, T.~E., Haberland, M., Reddy, T.,
  Cournapeau, D., Burovski, E., Peterson, P., Weckesser, W., Bright, J., {van
  der Walt}, S.~J., Brett, M., Wilson, J., Millman, K.~J., Mayorov, N., Nelson,
  A. R.~J., Jones, E., Kern, R., Larson, E., Carey, C.~J., Polat, {\.I}., Feng,
  Y., Moore, E.~W., {VanderPlas}, J., Laxalde, D., Perktold, J., Cimrman, R.,
  Henriksen, I., Quintero, E.~A., Harris, C.~R., Archibald, A.~M., Ribeiro,
  A.~H., Pedregosa, F., {van Mulbregt}, P., and {SciPy 1.0 Contributors},
  \enquote{{{SciPy} 1.0: Fundamental Algorithms for Scientific Computing in
  Python},} \emph{Nature Methods}, Vol.~17, 2020, pp. 261--272.
\newblock \doi{10.1038/s41592-019-0686-2}.

\bibitem[{Frank and Wolfe(1956)}]{frank1956algorithm}
Frank, M., and Wolfe, P., \enquote{An algorithm for quadratic programming,}
  \emph{Naval Research Logistics Quarterly}, Vol.~3, No. 1-2, 1956, pp.
  95--110.
\newblock \doi{https://doi.org/10.1002/nav.3800030109}.

\bibitem[{Jaggi(2013)}]{jaggi2013revisiting}
Jaggi, M., \enquote{Revisiting {Frank-Wolfe}: Projection-Free Sparse Convex
  Optimization,} \emph{Proceedings of the 30th International Conference on
  Machine Learning}, Proceedings of Machine Learning Research, Vol.~28, PMLR,
  2013, pp. 427--435.
\newblock \urlprefix\url{https://proceedings.mlr.press/v28/jaggi13.html}.

\bibitem[{Pedregosa et~al.(2020)Pedregosa, Negiar, Askari, and
  Jaggi}]{pedregosa2020linearly}
Pedregosa, F., Negiar, G., Askari, A., and Jaggi, M., \enquote{Linearly
  Convergent Frank-Wolfe with Backtracking Line-Search,} \emph{Proceedings of
  the Twenty Third International Conference on Artificial Intelligence and
  Statistics}, Proceedings of Machine Learning Research, Vol. 108, PMLR, 2020,
  pp. 1--10.
\newblock \urlprefix\url{https://proceedings.mlr.press/v108/pedregosa20a.html}.

\bibitem[{Lacoste-Julien and Jaggi(2015)}]{lacoste2015global}
Lacoste-Julien, S., and Jaggi, M., \enquote{On the Global Linear Convergence of
  Frank-Wolfe Optimization Variants,} \emph{Advances in Neural Information
  Processing Systems}, Vol.~28, Curran Associates, Inc., 2015.
\newblock
  \urlprefix\url{https://proceedings.neurips.cc/paper/2015/hash/c058f544c737782deacefa532d9add4c-Abstract.html}.

\bibitem[{Dvurechensky et~al.(2022)Dvurechensky, Safin, Shtern, and
  Staudigl}]{dvurechensky2022generalized}
Dvurechensky, P., Safin, K., Shtern, S., and Staudigl, M., \enquote{Generalized
  self-concordant analysis of Frank--Wolfe algorithms,} \emph{Mathematical
  Programming}, 2022, pp. 1--69.
\newblock \doi{10.1007/s10107-022-01771-1}.

\bibitem[{Liu et~al.(2016)Liu, Yuan, Zhang, Liu, and
  Metaxas}]{liu2016efficient}
Liu, B., Yuan, X.-T., Zhang, S., Liu, Q., and Metaxas, D.~N.,
  \enquote{Efficient k-support-norm regularized minimization via fully
  corrective frank-Wolfe method,} \emph{Proceedings of the Twenty-Fifth
  International Joint Conference on Artificial Intelligence}, AAAI Press, 2016,
  p. 1760–1766.
\newblock \urlprefix\url{https://www.ijcai.org/Abstract/16/252}.

\bibitem[{Gu{\'e}lat and Marcotte(1986)}]{guelat1986some}
Gu{\'e}lat, J., and Marcotte, P., \enquote{Some comments on Wolfe's ‘away
  step’,} \emph{Mathematical Programming}, Vol.~35, No.~1, 1986, pp.
  110--119.
\newblock \doi{10.1007/BF01589445}.

\bibitem[{Kuhn and Tucker(1951)}]{kuhn1951nonlinear}
Kuhn, H., and Tucker, A., \enquote{Nonlinear Programming,} \emph{Second
  Berkeley Symposium on Mathematical Statistics and Probability}, 1951, pp.
  481--492.

\bibitem[{Karush(1939)}]{karush1939minima}
Karush, W., \enquote{Minima of functions of several variables with inequalities
  as side constraints,} \emph{M. Sc. Dissertation. Dept. of Mathematics, Univ.
  of Chicago}, 1939.

\bibitem[{Bazaraa et~al.(2006)Bazaraa, Sherali, and
  Shetty}]{bazaraa2006nonlinear}
Bazaraa, M.~S., Sherali, H.~D., and Shetty, C.~M., \emph{The Fritz John and
  Karush–Kuhn–tucker Optimality Conditions}, John Wiley \& Sons, Ltd, 2006,
  Chap.~4, pp. 163--236.
\newblock \doi{10.1002/0471787779.ch4}.

\bibitem[{Beck(2017)}]{beck2017first}
Beck, A., \emph{First-Order Methods in Optimization}, Society for Industrial
  and Applied Mathematics, 2017, Chap.~6.
\newblock \doi{10.1137/1.9781611974997}.

\bibitem[{Beck and Teboulle(2009)}]{beck2009fast}
Beck, A., and Teboulle, M., \enquote{A Fast Iterative Shrinkage-Thresholding
  Algorithm for Linear Inverse Problems,} \emph{SIAM Journal on Imaging
  Sciences}, Vol.~2, No.~1, 2009, pp. 183--202.
\newblock \doi{10.1137/080716542}.

\bibitem[{Stark et~al.(1998)Stark, Yang, and Yang}]{stark1998vector}
Stark, H., Yang, Y., and Yang, Y., \emph{Vector space projections: a numerical
  approach to signal and image processing, neural nets, and optics}, John Wiley
  \& Sons, Inc., 1998, Chap.~3.

\bibitem[{Uriot et~al.(2021)Uriot, Izzo, Sim{\~o}es, Abay, Einecke, Rebhan,
  Martinez-Heras, Letizia, Siminski, and Merz}]{uriot2021spacecraft}
Uriot, T., Izzo, D., Sim{\~o}es, L.~F., Abay, R., Einecke, N., Rebhan, S.,
  Martinez-Heras, J., Letizia, F., Siminski, J., and Merz, K.,
  \enquote{Spacecraft Collision Avoidance Challenge: design and results of a
  machine learning competition,} \emph{Astrodynamics}, 2021, pp. 1--20.
\newblock \doi{10.1007/s42064-021-0101-5}.

\bibitem[{Vallado(2013)}]{vallado2001fundamentals}
Vallado, D.~A., \emph{Fundamentals of astrodynamics and applications},
  4\textsuperscript{th} ed., Microcosm Press, 2013, Chap.~10.
\newblock
  \urlprefix\url{https://microcosmpress.com/publishing/fundamentals-of-astrodynamics-and-applications-fourth-edition/}.

\bibitem[{Park et~al.(2019)Park, Stevenson, Nicolls, Lu, Griffith, and
  Rosner}]{park2019statistical}
Park, I., Stevenson, M., Nicolls, M., Lu, E., Griffith, N., and Rosner, C.,
  \enquote{Statistical covariance realism assessment of leolabs’ orbit
  determination system,} \emph{Advanced Maui Optical and Space Surveillance
  Technologies Conference}, 2019, p.~10.

\end{thebibliography}

\end{document}